\documentclass[fleqn,10pt]{wlscirep}
\usepackage{amsmath}
\usepackage{amsfonts}
\usepackage{amssymb}
\usepackage{enumerate}
\usepackage{ulem}
\usepackage{xcolor}
\usepackage{wasysym}
\usepackage{hyperref}
\usepackage{mathtools}
\usepackage{float}
\usepackage{graphicx}
\usepackage[font={small}]{caption}

\newcommand{\bb}[1]{\textcolor{black}{#1}}

\definecolor{OliveGreen}{rgb}{0,0.6,0}

\newcommand{\bra}[1]{\left[\, #1 \, \right]}

\newcommand{\pare}[1]{\left(\, #1 \, \right)}

\renewcommand{\t}[1]{{\tilde{#1}}}

\title{A biophysical model explains the spontaneous bursting behavior in the developing retina }
\author[1]{Dora Matzakou-Karvouniari}
\author[2]{Lionel Gil}
\author[3]{Elaine Orendorff}
\author[3]{Olivier Marre}
\author[3]{Serge Picaud}
\author[1*]{Bruno Cessac}
\affil[1]{Biovision team, Université Côte d'Azur, Inria, France}
\affil[2]{INPHYNI, Université Côte d’Azur, CNRS, France}
\affil[3]{Institut de la Vision, Paris, France}

\affil[*]{bruno.cessac@inria.fr}

\graphicspath{{./Figures/}}

\begin{abstract}
During early development, waves of activity propagate across the retina and play a key role in the proper wiring of the early visual system. During a particular phase of the retina development (stage II) these waves are triggered by a transient network of neurons, called Starburst Amacrine Cells (SACs), showing a bursting activity which disappears upon further maturation. The underlying mechanisms of the spontaneous bursting and the transient excitability of immature SACs are not completely clear yet. While several models have attempted to reproduce retinal waves, none of them is able to mimic the rhythmic autonomous bursting of individual SACs and reveal how these cells change their intrinsic properties during development. Here, we introduce a mathematical model, grounded on biophysics, which enables us to reproduce the bursting activity of SACs and to propose a plausible, generic and robust, mechanism that generates it. The core parameters controlling repetitive firing are fast depolarizing $V$-gated calcium channels and hyperpolarizing $V$-gated potassium channels. The quiescent phase of bursting is controlled by a slow after hyperpolarization (sAHP), mediated by 
calcium-dependent potassium channels. Based on a bifurcation analysis we show how biophysical parameters, regulating calcium and potassium activity, control the spontaneously occurring fast oscillatory activity followed by long refractory periods in individual SACs. We make a testable experimental prediction on the role of voltage-dependent potassium channels on the excitability properties of SACs and on the evolution of this excitability along development. We also propose an explanation on how SACs can exhibit a large variability in their bursting periods, as observed experimentally within a SACs network as well as across different species, yet based on a simple, unique, mechanism. As we discuss, these observations at the cellular level have a deep impact on the retinal waves description.

\end{abstract}

\begin{document}
\flushbottom
\maketitle

\thispagestyle{empty}

\section*{Introduction}

Retinal waves, observed in many vertebrate species - chicks \cite{sernagor-eglen-etal:00}, ferrets \cite{feller-butts-etal:97}, mice \cite{maccione-hennig-etal:14}, turtles \cite{sernagor-grzywacz:99}, macaques \cite{warland-huberman-etal:06} etc. are spontaneous bursts of activity propagating in the developing retina and playing a fundamental role in shaping the visual system and retinal circuitry. They emerge due to the conjunction of intrinsic single-cell properties (excitability and long refractoriness) and network interactions \cite{zheng-lee-etal:06}. In the developing retina, waves evolve during three consecutive stages, mainly characterized by different types of synaptic transmission; gap junctions (stage I), acetylcholine (stage II) and glutamate (stage III) \cite{sernagor-hennig:12} and are mediated by transient networks of specific cell types.
An important type of such cells are the Starburst Amacrine Cells (SACs), which are involved in direction selectivity in the mature retina \cite{fried-munch-etal:02,yoshida-watanabe:01} thanks to their special asymmetric synaptic distribution of excitatory and inhibitory synapses. However, during early development, these cells have a different role; they are responsible for eliciting stage II retinal waves via their spontaneous bursting behavior induced by autonomous intrinsic cellular mechanisms \cite{zheng-lee-etal:06}. This activity disappears completely upon maturation \cite{zheng-lee-etal:06}. During stage II, SACs constitute a transient network of autonomous bursters which, through cholinergic coupling, achieve opportunistic local synchrony. Under conditions yet to be identified, this local area of synchronized bursters can give rise to propagating waves.
Although stage II retinal waves are quite thoroughly studied experimentally \cite{feller-butts-etal:97,sernagor-eglen-etal:00, maccione-hennig-etal:14} as well as theoretically by several modeling approaches \cite{godfrey-swindale:07,hennig-adams-etal:09,lansdell-ford-etal:14,godfrey-eglen:09} (see Methods for a short review of these models and a comparison to ours) there are still important questions not yet addressed by existing studies. So far, previous models do not allow to understand how biophysical parameters, like conductances, impact the activity of SACs during this stage. This question is important as some of these parameters evolve during development, or can be controlled pharmacologically, with drastic effects on individual SACs bursting \cite{zheng-lee-etal:04,zheng-lee-etal:06}. Finally, and although stage II retinal waves are ubiquitous, experiments reveal a vast variability in the bursting period of SACs across species. Are SACs fundamentally different from one species to another, or is there a common mechanism of bursting under which the period is quite sensitive to physiological parameters variations ?  

In the spirit of previous works\cite{han-tanimura-etal:17,park-rubin:06,choi-zhang-etal:14} using dynamical systems and bifurcation theory to characterize neural systems, we develop a thorough modeling of the SACs bursting which accurately reproduces the experimental observations made in the seminal papers  by Zheng et al. \cite{zheng-lee-etal:04, zheng-lee-etal:06}, several of them for the first time. In addition to new theoretical predictions, we also propose a simple scenario for the SACs bursting mechanism and its evolution during development, that could be tested experimentally. Bursting is thoroughly studied in the neuronal modeling literature \cite{izhikevich:07}, as well as the possible bifurcations scenarios generating it. Through our analysis, we identify which parameters are essential to control the transient bursting of immature SACs occurring naturally along development or pharmacologically. As parameters variation is restricted to small ranges by biophysical constraints, our hypothesis is that, during development, SACs are close to a bifurcation point, consequently making them sensitive to small physiological parameters variations.

More precisely, we propose to describe and explain the experimental observations  of Zheng et al.\cite{zheng-lee-etal:04, zheng-lee-etal:06} on individual immature SACs dynamics, in the following theoretical framework: SACs burst because, during stage II, they are close to a (saddle-node) bifurcation, driving the cell from a rest state to fast oscillations. The bifurcation is triggered by  small variations of the membrane potential, due to noise or other cellular effect. 
After this bifurcation, SACs stay in the regime of fast repetitive firing, triggering the activation of slow calcium-gated potassium channels generating a slow hyperpolarization current (sAHP). This current eventually drives the cell to a homoclinic bifurcation. As a consequence, cells return back to the quiescent state allowing a long after hyperpolarization phase.
This bifurcation analysis reveals a mechanism for SACs rhythmic bursting, interpretable with respect to biophysics and the experiments of Zheng et al.\cite{zheng-lee-etal:04, zheng-lee-etal:06} on individual immature SACs dynamics.
 In addition, we identify the fast voltage-gated potassium channels as a possible candidate for the fast hyperpolarizing channels involved in bursting (to our best knowledge not yet been experimentally determined). Based on \cite{ozaita-petit-jacques-etal:04,kaneda-ito-etal:07} we propose they could be of $Kv_3$ type. This leads us to a conjecture on the role of the $Kv_3$ potassium conductance variation which may act on the transient intrinsic properties of SACs excitability during development. Furthermore, the closeness to a bifurcation has strong consequences on the interburst intervals of SACs. It is naturally associated with a high variability which can explain the wide range of interburst periods observed within a network \cite{ford-feller:12}, as well as across species \cite{godfrey-eglen:09}. As we show, a small external current ($\sim 5$ pA) strongly impacts the interburst intervals of SACs. This implies, that, in a network of cholinergically coupled SACs, the bursting period has a wide intrinsic variability, resulting from simple bifurcation arguments, without requiring exogenous mechanisms. In the discussion section, we briefly argue how a set of such inhomogeneous bursters may achieve synchrony and wave generations without any additional mechanism than the existing cholinergic interaction. The general discussion on how cholinergic interaction between nearby autonomous bursters generates wave-like, population-level behavior propagating in a similar way as is observed experimentally will be addressed in a forthcoming paper (see \cite{karvouniari-gil-etal:16,karvouniari-gil-etal:17,karvouniari:18}). \\
The paper is organized as follows. The section 'Results' introduces the model definition, the bifurcation analysis, and the main conclusions resulting from this analysis, including experimental predictions. The section 'Discussion' is mainly dedicated to the biological interpretation of our theoretical results, as well as their impact on the description of retinal waves. Finally, in Methods, we provide the technical aspects of our work including the biophysical derivation of the model's equations, as well as mathematical computations used in the paper. We also include in this section a short review on the comparison of existing models.

\section*{Results}

\bb{During retinal development, Starburst Amacrine Cells (SACs), which are cells forming a lattice right above the ganglion cells layer, are found to exhibit spontaneous intrinsic rhythmic bursting activity, which disappears upon maturation \cite{zheng-lee-etal:06}. In the following, we derive a biophysical set of equations, describing the bursting activity of immature SACs, following experimental observations regarding the ionic currents involved in their bursting \cite{zheng-lee-etal:06}. We calibrate the equations parameters in order to reproduce the bursting characteristics found in \cite{zheng-lee-etal:04,zheng-lee-etal:06}. The analysis of our model leads to our central hypothesis, that the rest state of SACs lies near a bifurcation point, which means that small variations of parameters could lead to abrupt changes in the dynamics of the cell. This hypothesis, which is novel for immature SACs, jointly explains A) the biophysical mechanism of bursting in the developing SACs, B) accounts for the vast variability of bursting characteristics of SACs across species and C) predicts a possible mechanism on how SACs lose their excitability upon maturation. \\
Let us now start by deriving our model along with the biological justifications for our equations.}

\subsection*{A biophysical model for bursting immature SACs}

\bb{In order to address the mechanism of bursting and how the physiological parameters of SACs change during development we need to model their underlying cellular mechanisms.}
In the case of immature SACs, the two key biophysical mechanisms associated with the emergence of spontaneous bursting during early development are \cite{zheng-lee-etal:06}:

\begin{itemize}
\item (i) fast repetitive bursts of spikes mainly controlled by fast voltage-gated channels;
\item (ii) prolonged After HyperPolarizations (AHPs) modulating fast oscillations, controlled by $Ca^{2+}$-gated $K^+$ channels.
\end{itemize}

\bb{In order to account for (i) and (ii), we model immature SACs activity with a conductance based model of the Morris-Lecar type \cite{morris-lecar:81}, an electronic-circuit-equivalent equation for the local membrane potential, with additional non-linear currents. Our model involves $5$ variables, whose evolution is controlled by a set of non-linear differential equations with $3$ timescales: (1) fast variables ($\sim 10 ms$) $V(t)$, the local membrane potential, and, $N(t)$, the gating variable for fast voltage-gated $K^+$ channels; (2) medium variable ($\sim 2$ s) $C(t)$, the intracellular $Ca^{2+}$ concentration ; (3) slow variables ($\sim 10$ s) $R(t)$ and $S(t)$, the gating variables for slow $Ca^{2+}$-gated $K^+$ channels. All parameters values and the auxiliary functions involved are found in Methods.}

\bb{The membrane voltage $V(t)$ obeys:
\begin{equation} \label{eq:Voltage1}
C_{m}\frac{d V}{d t}=I_L(V)+I_{C}(V)+I_K(V,N)+\underbrace{I_{sAHP}(V,R)  + I_{ext}}_{I_{tot}} + \sigma \xi_t,
\end{equation}
where $C_m$ is the membrane capacitance, $I_L=-g_L(V-V_L)$  is the leak current, with $g_L$, leak conductance and $V_L$, leak reversal potential.\\}

\bb{The fast repetitive firing (see (i)) during the active phase of bursting generally results from the competition between a depolarizing and a hyperpolarizing current. Experiments in \cite{zheng-lee-etal:06} have specified voltage-gated $Ca^{2+}$ channels as the depolarizing component of bursting in SACs. However, the ionic channels related to the fast hyperpolarizing component of SACs bursting have not yet been identified experimentally. In this work, we propose that \textit{fast} voltage-gated $K^+$ channels play this role (see also Discussion for further justification). Therefore, the terms $I_C$ and $I_K$, respectively corresponding to $Ca^{2+}$ and $K^+$ currents, are generating the fast oscillations in our model. \\
Note that \cite{zheng-lee-etal:06} have shown that voltage-gated $Na^+$ channels do not participate in the bursting mechanism of immature SACs (bursting activity of SACs was not altered upon tetrodoxin -TTX- application), thus, dynamics of $Na^+$ channels will not be considered in our modeling (see also Discussion).}\\

\bb{The voltage-gated $Ca^{2+}$ current is described by a Morris-Lecar model \cite{morris-lecar:81} where:
\begin{equation} \label{eq:ICa}
I_{C}(V)=-{g}_{C} M_\infty(V)(V-V_{C}).
\end{equation}
${g}_{C} M_\infty(V)$ is the voltage dependent conductance of the $Ca^{2+}$ channel (see Eq \eqref{eq:gcav} in Methods).\\
The fast voltage-gated $K^+$ channel is modeled as:
\begin{equation}\label{eq:IK}
I_K(V,N)=-g_{K}N(V-V_{K}).
\end{equation}
where the evolution of the fast voltage-gated $K^+$ channel gating variable $N(t)$ is given by:
\begin{equation}\label{eq:N}
\tau_N\frac{d N}{d t}=\Lambda(V)(N_{\infty}(V)-N),
\end{equation}
$\Lambda(V)$ and $N_{\infty}(V)$ are given by Eq \eqref{eq:Lambda}, \eqref{eq:Ninf} in Methods.\\
The form of equations \eqref{eq:IK}, \eqref{eq:N} is based on \cite{morris-lecar:81}. However, the timescale of the variable $N$ is much faster ($\sim ms$) in our model that in the original Morris-Lecar model ($\sim s$), calibrated in order to capture the frequency of the fast repetitive firing of SACs, which is around $20$ Hz \cite{zheng-lee-etal:06} (see also Bifurcation Analysis).}

\begin{figure}[h!]
\centering\includegraphics[width=\linewidth,trim={0 10cm 0 0},clip]{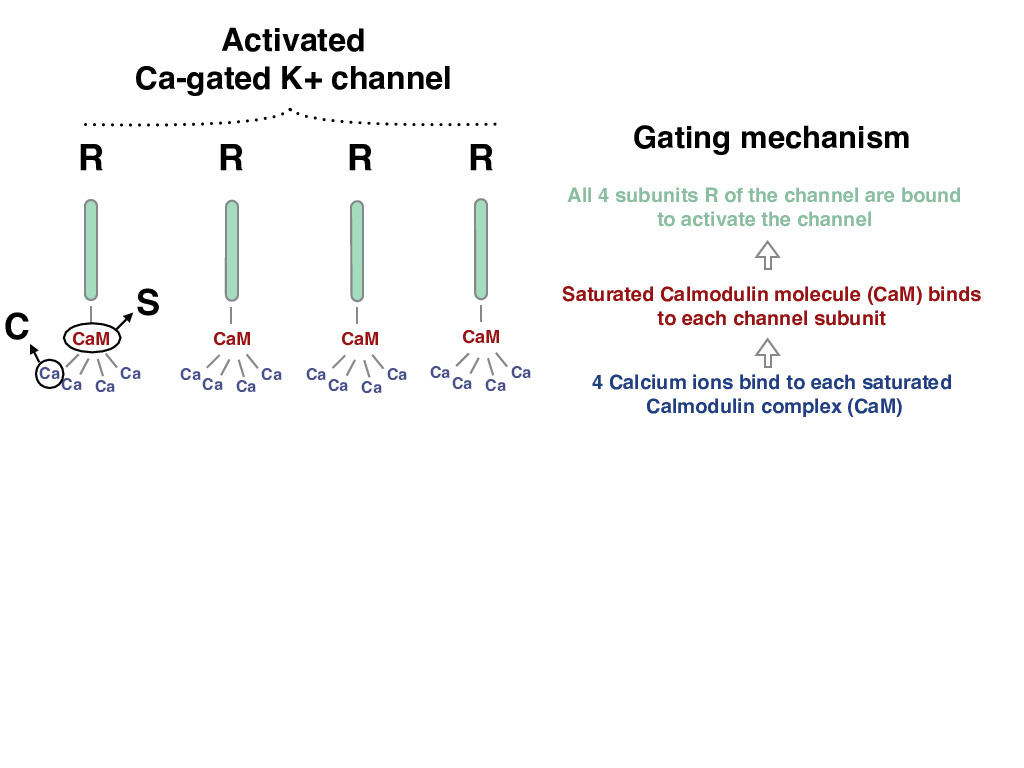}
\caption{\label{Fig:Scehmasahp} \textbf{Schematic representation of the modeling of the gating mechanism of $Ca^{2+}$-gated $K^+$ channels. The correspondence between the channel's activation steps and the modeling state variables $R,S,C$ is also indicated.} }
\end{figure}

The long refractoriness in-between consecutive bursts is controlled by a slow After Hyper-Polarization (sAHP) $K^+$ current, $I_{sAHP}$. It was observed by \cite{zheng-lee-etal:06} that $I_{sAHP}$ is mediated by $Ca^{2+}$-gated $K^+$ channels, and that it resembles the sAHP observed by Abel et al. \cite{abel-lee-etal:04}, generated by specific channels called $SK$. Following these tracks we propose a modeling of "SK"-like channel (as named in \cite{zheng-lee-etal:06}) based on \cite{hennig-adams-etal:09} for the structure of the equations and \cite{graupner-erler-etal:05} for the calcium dynamics. The mechanism of the opening of $Ca^{2+}$-gated ionic channels channels is analyzed in detail in the "Methods" section. In order to simplify the cascade of chemical reactions taking place while opening the sAHP channels, we approximate the channel dynamics by reducing the process into two discrete steps: a) Four ions of $Ca^{2+}$ bind to a second messenger protein called calmodulin, forming a saturated calmodulin complex, CaM; b) CaM binds to each of the four intracellular subunits of the channel to open it (see Fig \ref{Fig:Scehmasahp}). This process is mapped to our model through three variables: 1) the variable $C$ which models the intracellular calcium concentration and mainly controls the gating variables of the sAHP channels, 2) the variable $S$ which models the fraction of the saturated calmodulin and 3) the variable $R$ which models the fraction of bounded terminals.  This gating mechanism is sketched in Fig \ref{Fig:Scehmasahp}.
\\
 
The sAHP current takes the form:
\begin{equation}\label{eq:IsAHP}
I_{sAHP}(V,R)=-g_{sAHP}R^4(V-V_{K}),
\end{equation}
where $g_{sAHP}$ is the maximum sAHP conductance. Indeed, $4$ bound terminals are needed to open a $Ca^{2+}$-gated $K^+$ channel, thus the corresponding conductance is $g_{sAHP}R^4$, involving a fourth order nonlinearity. 

We model the gating mechanism of the $Ca^{2+}$ gated $K^+$ channels  as follows.
The gating variable $R(t)$ obeys:
\begin{equation}\label{eq:R}
\tau_{R}\frac{d R}{d t}= \alpha_R \, S(1-R)-R,
\end{equation}
the fraction of saturated calmodulin concentration $S(t)$ obeys:
\begin{equation}\label{eq:S}
\tau_S \frac{dS}{dt} = \alpha_S C^4 (1-S) - S,
\end{equation}
and the intracellular calcium concentration $C(t)$ obeys: 
\begin{equation}\label{eq:Ca}
\tau_{C}\frac{d C}{d t}=-\frac{\alpha_{C}}{H_{X}}C+C_{0} + \delta_{C} \, I_C(V).
\end{equation} 
The derivation of the equations \eqref{eq:R}, \eqref{eq:S}, \eqref{eq:Ca} is fully justified in Methods.\\

Finally, $\xi_t$ is a white noise whose amplitude is constant with time and controlled by $\sigma$ and $I_{ext}$ is an external current. 
For subsequent analysis we introduce the current $I_{tot}=I_{sAHP}  + I_{ext}$.

\paragraph{Dynamical changes with respect to parameters variations.} 
In our model, the variable $V, N$ evolve with a fast time scale of the order of a few milliseconds under the influence of the current $I_{sAHP}$  whose conductance is slow, being driven by slow variables $C,S,R$  (time scale - several seconds).
As a consequence, the main material of this paper relies 
on a thorough analysis of the bifurcation structure in the fast Morris-Lecar dynamics. There already exist remarkable numerical bifurcation analyses of the Morris-Lecar model (see for example the very detailed work of \cite{tsumoto-kitajimac-etal:06}) which has however many parameters: changing their range of variations can dramatically impact the dynamics. Our study on SACs bursting considers quite different range of parameters than the studies we know about (more related to the original paper investigating the barnacle giant muscle \cite{morris-lecar:81}). For example, \cite{tsumoto-kitajimac-etal:06} investigates the effect of a positive external current whereas we consider a negative external current, with quite different effects on dynamics. Other parameters such as $V_1,V_2,V_3,V_4, V_L, g_K, g_{Ca}$ have very different ranges in our case (e.g. $V_3=-1$ mV in \cite{morris-lecar:81}, $V_3 \in [0,25]$ mV in \cite{tsumoto-kitajimac-etal:06}, $V_3=- 25 mV$ in our case).   
Finally, the characteristic rate for opening potassium channels (variable $N$) is of order $\frac{1}{15}$ $s^{-1}$ in Morris-Lecar's paper, and of order $0.1$ $s^{-1}$ in \cite{tsumoto-kitajimac-etal:06}. That is the variable $N$ is slow. In contrast, our variable $N$, mimicking fast potassium channels is very fast ($\frac{1}{\tau_N}=200$ $s^{-1}$). This results in a very fast contraction in the $N$ direction near the rest state visible e.g. in the shape of stable and unstable manifolds (see Fig \ref{Fig:BifurcationDiagram_Iext_gK_VL-70} A,B;  \ref{Fig:BifurcationDiagram_gCa_gK_VL-70} B,C; \ref{Fig:BifurcationDiagram_gCa_gK_VL-72} B,E,G; \ref{fig:HeatMapIBI_vs_gK_V3_theta1_sigma4_VL-72} c,d). As a consequence, our bifurcations study is new and, in addition to explain bursting in SACs, brings original results in the analysis of the Morris-Lecar model dynamics.

We distinguished two classes of parameters. The first one consists of modeling parameters constraining the Morris-Lecar dynamics, namely $V_1,V_2,V_3,V_4$ (see Methods for the corresponding equations). We verified the robustness of the scenario described in the paper when varying these parameters. Especially, $V_3$ is the half activation potential of the fast potassium channels, playing a central role in SACs excitability as shown later in the text. 
The second class is the set of parameters constrained by biophysics such as conductances, reversal potentials, capacitance, ... Most of them have been fixed based on the biophysical literature and are not considered to vary. Here, we choose to study the variation of $4$ of these parameters: $I_{tot}$, because it controls bursting, and $g_K, g_C$ and $V_3$ because, as we show later, their variations allows us to reproduce experimental facts, that have not been reproduced before by any model. The most striking consequence is a conjecture on the role of the potassium conductance during development.

\subsection*{Bifurcation analysis} \label{sec:Dyn-Analysis}
\bb{Bursting is an alternation between a rest state and repetitive firing, often modulated by slow voltage or $Ca^{2+}$ dependent processes \cite{izhikevich:07}. In order to address which are the biophysical mechanisms for a) triggering, and, b) sustaining bursting in immature SACs, we will perform bifurcations analysis.  This is a powerful mathematical tool providing a compact, geometric representation, of dynamical systems behavior when parameters are modified. While, in general, a continuous variation of a parameter induces a continuous variation of the dynamical system's solutions, the variation of parameters around bifurcation points induces, in contrast, strong and abrupt changes in the dynamics.}

\begin{figure}[!htt]
\centerline{
\includegraphics[width=\linewidth]{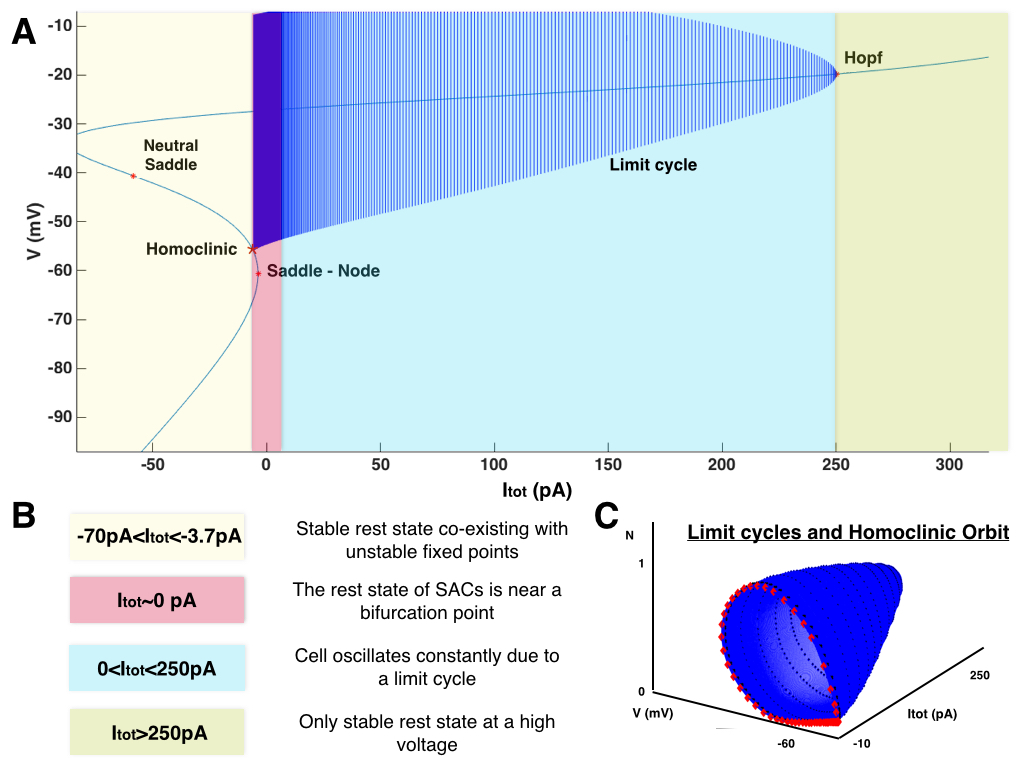}
}
 \caption{\label{Fig:BigDiagML} \bb{\textbf{A. Bifurcation diagram of the fast dynamics ($N, V$)} (Eq \eqref{eq:ML_fast} in Methods) \textbf{when the constant current $I_{tot}$ is varied in a wide range.} Each deep blue curve corresponds to the projection of a trajectory of the fast subsystem onto the $V$ axis, for a fixed value of $I_{tot}$. \textbf{B.} Interpretation of all the different dynamical regimes of the bifurcation diagram. \textbf{C.} Same as in A, except that the various trajectories are now drawn in the $(N,V)$ plane and not only projected onto the $V$ axis. \textit{Blue.} Evolution of the limit cycles when $I_{tot}$ varies. \textit{Red.} Homoclinic orbit. Bursting stops by an homoclinisation of the limit cycle.}} 
\end{figure}

\bb{We first perform a bifurcation analysis of the fast $V,N$ dynamics in the presence of the current $I_{tot}$, constant in time, and used as a bifurcation parameter. The bifurcation diagram is represented in Fig \ref{Fig:BigDiagML} A (drawn using MATCONT \cite{dhooge-govaerts-etal:03}), where we have explored a wide range of variation of $I_{tot}$ ($-70$ to $+310$ pA). For the analysis, we consider an example with the potassium $g_K=10$ nS, and calcium conductance $g_{C}=12$ nS (see Robustness with respect to parameters variations). We reveal a rich dynamical profile upon the variation of an external current where for $I_{tot}<-3.7 pA$, stable rest state co-exists with unstable fixed points, for $-3.7<I_{tot}<250 pA$, cell oscillates constantly due to a limit cycle and for $I_{tot}>250 pA$ only a stable rest state at a high voltage exists (see Fig \ref{Fig:BigDiagML} B). We refer the reader to e.g. \cite{guckenheimer-holmes:83} for the dynamical systems terminology used in the next paragraphs.}

\bb{
In more detail, when $I_{tot} \lessapprox -5.8$ pA, there is a stable rest state with a membrane potential $V_{rest} \in [-70,-60]$ mV, coexisting with two unstable fixed points, a saddle and an unstable focus (see Fig 9A for the phase portrait of this case). When $I_{tot} \simeq -5.8 pA = I_{H_c}$, there is an homoclinic bifurcation giving rise to a stable limit cycle. In the range $I_{tot} \in [I_{H_c}, I_{SN_1}]$, where $I_{SN_1} \simeq -3.7$ pA, the stable rest state coexists then with a stable limit cycle, corresponding to fast oscillations, and an unstable fixed point. When $I_{tot} = I_{SN_1} \simeq -3.7$ pA the stable rest state coalesces with the middle unstable branch and both disappear with a saddle-node bifurcation ($SN_1$).  For $I_{tot} > I_{SN_1}$ the dynamics has only one attractor, the limit cycle, corresponding to fast oscillations. This cycle eventually disappears by a Hopf bifurcation at  $I_{tot}=250$ pA (Fig \ref{Fig:BigDiagML}). However, this value of external current is quite beyond the range of plausible values during the bursting activity of SACs.}

\bb{
We observe that the rest state of SACs, corresponding to $I_{tot}=0$ pA, lies near a saddle-node bifurcation point within a narrow region of a few $pA$, which is our main emerging hypothesis. This leads to important consequences for SACs described in the following.}

\begin{figure}[ht!]
\centering
\includegraphics[width=\linewidth,height=8cm]{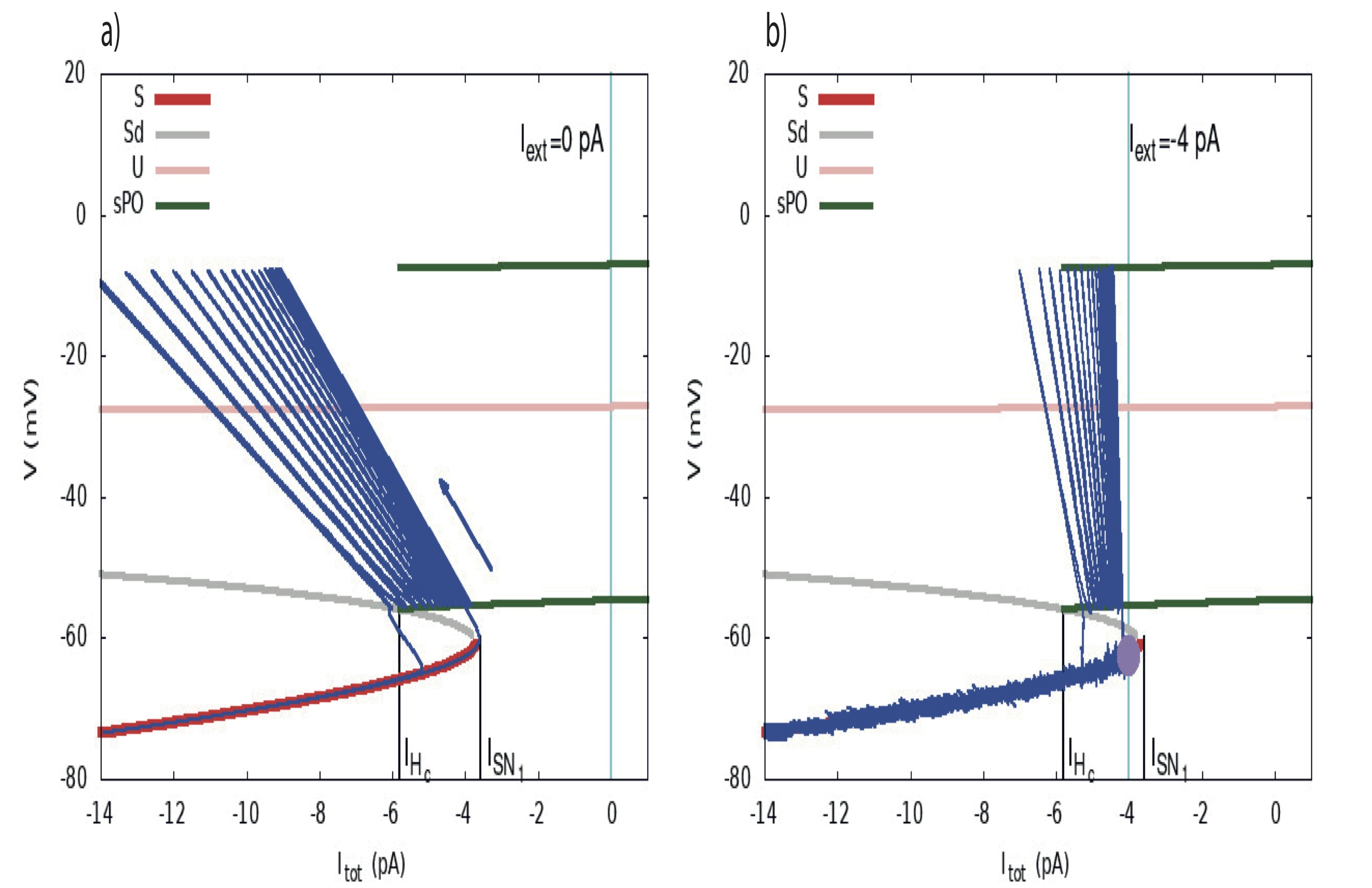}
 \caption{\label{Fig:SpontaneousBursting}
 \textbf{The two scenarios of triggering bursting. a) Dynamically driven bursting,  $I_{ext}=0$}. The plot shows a representation of bursting in the plane $(I_{tot},V)$ (zoom of the bifurcation diagram of Fig \ref{Fig:BigDiagML}, near $I_{SN}$). Red lines correspond to stable fixed points, gray and pink lines to unstable fixed points, green lines are the extremal values of voltage fast oscillations. Upon this bifurcation diagram, obtained from the reduced model (Morris-Lecar with $I_{sAHP}=I_{tot}$ = constant), we superimpose the trajectory of the full system in the plane $(I_{sAHP},V)$ (deep blue). The blue arrow indicates the direction of the flow. In the fast oscillations regime $V$ is varying periodically, with a fast period, inducing a fast variation of the term $V-V_K$ in the sAHP current, explaining the diagonal motion. Despite these fast oscillations one sees that the bifurcation is driven by the slow motion on the branches obtained from the bifurcation analysis assuming a constant current. The conductance $g_{sAHP}R^4$ is varying slowly, with the time scale of $R$, explaining the slow left-wise shift of the trajectory until the homoclinic bifurcation point is reached. The cyan vertical line corresponds to the value of $I_{ext}=0$. \textbf{b) Noise driven bursting.} The dynamics has now a stable fixed point (purple circle). An additional current, $I_{ext}=-4$ pA (vertical cyan line) is now present. Bursting cannot occur without an exogenous excitation, here noise. The amount of noise (controlled by $\sigma$) necessary to induce bursting depends on the distance between the fixed point (controlled by $I_{ext}$) and the bifurcation point $I_{SN_1}$.}
\end{figure}

\paragraph{Bifurcations analysis reveals two possible drives for triggering bursting.}
\bb{The bifurcations analysis for the fast subsystem shown in Fig \ref{Fig:BigDiagML}, reveals all possible dynamical regimes upon the variation of a constant external current within a wide interval of our choice. Thanks to the timescales separation in our model (fast, medium, slow), we make the approximation that the bifurcation diagram in Fig \ref{Fig:BigDiagML}A, does not change significantly when we consider the full model case. However, in the full model $I_{tot}$ is not constant, but time-dependent, driven by sAHP. By superimposing the solution of the full model and the bifurcation diagram of the fast subsystem, we reveal how the trajectory driven by the slow variations of $I_{sAHP}$ matches the bifurcation diagram, in the plane $(I_{tot}, V)$. See Fig \ref{Fig:SpontaneousBursting} showing a zoom of the bifurcation diagram shown in Fig \ref{Fig:BigDiagML}A, to our regime of interest (\textit{stable branch}: red, \textit{neutral saddles}:  gray, \textit{unstable branch}: pink, green traces showing the maximum and minimum of $V$ along the limit cycle), with the superimposed trajectories of the full system (blue curves).}\\

\bb{Analyzing how the slow variations of $I_{sAHP}$ generate bursting, we actually exhibit two possible drives of triggering bursting: 
}

\bb{i) \textit{Dynamical drive}, induced by the mere dynamics of neurons, without external influence.
In our model, the joint fast dynamics of $Ca^{2+}$ and $K^+$ channels generates fast oscillations while the slow AHP, mediated by $Ca^{2+}$ gated $K^+$ channels, modulates slow oscillations \cite{zheng-lee-etal:06}. The conjunction of these two mechanisms generates bursting. Note that in this case, $I_{ext}=0$ in Eq \eqref{eq:Voltage1}, meaning there is no external current (Fig \ref{Fig:SpontaneousBursting} left). In this scenario, cells burst periodically, with a frequency controlled by the characteristic times $\tau_R, \tau_S$ of variables $R$ and $S$ respectively. From a dynamical systems perspective, there are two important bifurcations associated with bursting: a) bifurcation of the rest state that leads repetitive firing and b) bifurcation of a spiking attractor that leads to a rest state \cite{izhikevich:07}. There are two possible types of bifurcations linked to such type of behavior (1) Homoclinic saddle-node bifurcation and (2) Hopf bifurcation. We have therefore shown that (1) holds in our case. The first bifurcation associated with the breaking of the rest state is the saddle-node bifurcation. The "spiking attractor", here the limit cycle, is created by an homoclinisation. According to the classification of bursting made by Izhikevich in \cite{izhikevich:07}, this corresponds to a "square-wave" point-cycle planar burster.}\\ 

\bb{ii) \textit{Noise-drive}, induced by an external excitation, typically noise. In order to move to this regime, it suffices to add an external constant current $ I_{ext}<I_{SN_1}=-3.7 pA$ to Eq \eqref{eq:Voltage1}. Then, the \textit{deterministic} dynamics converges towards a stable fixed point, the cell ends up in a steady rest state and cannot switch to rapid firing. In this case, the bifurcation diagram (Fig \ref{Fig:SpontaneousBursting} right) has now two fixed points: $V_{rest}$, which is stable (purple point), and $V_u$ which is unstable (intersection of the gray branch and the vertical blue line). However, in the \textit{presence of noise}, sufficiently close to the bifurcation point (depending on the noise amplitude), the random fluctuations around the rest state leads eventually the cell into the rapid firing regime (Fig \ref{Fig:SpontaneousBursting} right). Note that when $I_{ext}$ is sufficiently large in absolute value the total current $I_{tot}$ is smaller than $I_{SN_1}$ whatever the (negative) value of $I_{sAHP}$. \\
}
\bb{
Taken together, since noise is always present in neurons, distinguishing those two scenarios for triggering bursting is not easy experimentally but still possible (see Discussion). Maybe both scenarios are possible in the same network of SACs, since it would correspond to a slight variability in the intrinsic parameters which control the rest state of each SAC, placing it on the right or left of the bifurcation within a narrow region ($\sim 10 pA$ according to our model). We are actually able to capture both the dynamically and the noise driven bursting regimes in a single, unifying model. In the presence of noise the dynamically driven bursting regime still exists but the period has now random fluctuations. In fact, the noise smooths the transition between those $2$ regimes. Also, what our analysis reveals is that, in the noise driven regime, bursting will occur only if the noise has a sufficiently large amplitude compared to the distance between the rest state and the bifurcation point. On the other hand, in the dynamically driven regime, bursting will occur only if the rest state is sufficiently close to the bifurcation in order for the amplitude of the negative sAHP current to be sufficient to compensate the depolarizing current ($I_{C}$) and enter the hyperpolarization phase. Otherwise, if the rest state is further on right from the bifurcation, SACs would exhibit only a repetitive firing regime without hyperpolarization. In other words, in both regimes, \textit{the cell has to be close enough to the bifurcation point to observe bursting.}}\\


\paragraph{Bifurcations analysis reveals a biophysical mechanism for SACs sustained bursting}

\bb{Our analysis reveals also a possible biophysical mechanism to explain how bursting activity is sustained periodically in immature SACs. Once fast repetitive firing is elicited in SACs, either through i) dynamical, or ii) noise drive, the biophysical cyclic mechanism which produces bursting is the \textit{same}. From the bifurcation analysis we see that once the cell crosses the saddle-node $SN_1$, it is firing rapidly (limit cycle). Being in a high voltage state, $Ca^{2+}$ loads, leading to a slow rise of sAHP conductance. This corresponds, in the bifurcation diagram, to a motion towards negative current values, with a slow variation due to the sAHP conductance driven by the slow variable $R$, and fast oscillations due the fast variable $V-V_K$ appearing in the sAHP current $I_{sAHP}=-g_{sAHP}R^4(V-V_K)$. As revealed by a bifurcation analysis where the sAHP conductance is the bifurcation parameter (not shown), and as shown in figure \ref{Fig:SpontaneousBursting}, these fast oscillations have no effect on the bifurcation, what matters is the slow variation. The increase of sAHP conductance continues until the cell reaches the homoclinic bifurcation where rapid firing stop. $I_{tot}=I_{sAHP}$ is very negative now, leading to the hyperpolarization phase. Then, because voltage is low, $[Ca^{2+}]$ decreases and $I_{sAHP}$ drops down. Eventually, SAC crosses the $SN_1$ bifurcation point and starts firing again either due to dynamics in one case or noise. This is illustrated in Fig \ref{Fig:SpontaneousBursting}. The deep blue traces are the trajectory of a burst in the plane $(I_{tot}=I_{sAHP},V)$, for the full system. One sees in particular the fast oscillations coming from the fast variable $V-V_K$.\\ }

\begin{figure}[h!]
\centering
\includegraphics[width=\linewidth]{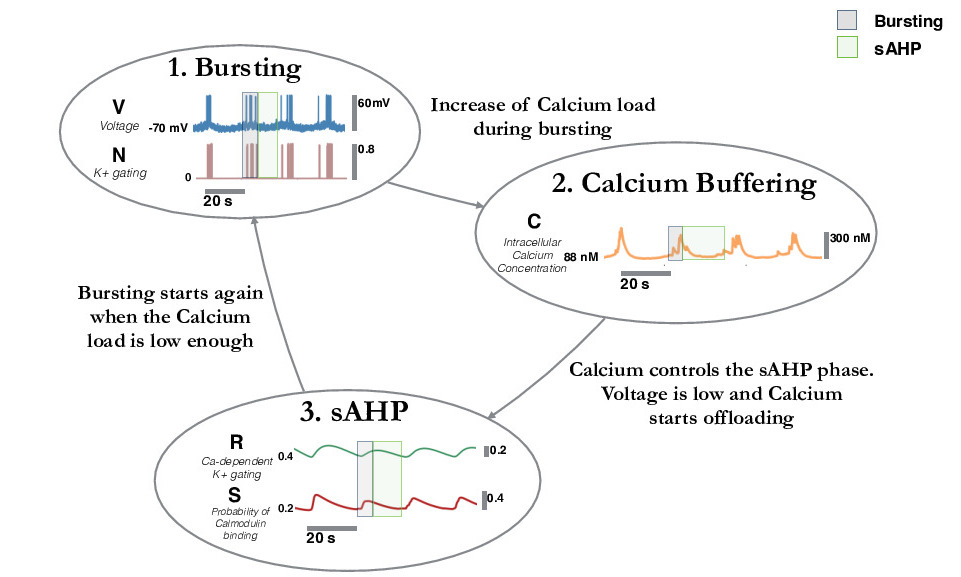}
\caption{\label{Fig:ScehmaModel} \textbf{Biophysical mechanism for bursting in immature SACs.} 1. Fast spiking occurs due to the competition between voltage gated $Ca^{2+}$ (excitatory current) and $K^+$ channels (inhibitory current). 2. Calcium load increases during the rapid firing phase while the voltage is in a high voltage state. This leads to a slow increase of sAHP. When sAHP is large enough there is a sharp decay of the voltage (bifurcation) and the hyperpolarisation phase of the cell starts. As voltage is low, the calcium load starts to slowly decrease. 3. During the slow offloading calcium stage we observe a refractory phase. The decrease of calcium concentration induces a slow decay of the sAHP. When calcium is small enough rapid firing starts again via a new bifurcation.}
\end{figure}

\bb{Nevertheless, for \textit{both} proposed scenarios the biophysical mechanism for sustaining bursting is the same. Although the conclusion drawn on the model is entirely based on a mathematical analysis via bifurcation theory, we have sketched the main mechanism of SACs bursting in Fig \ref{Fig:ScehmaModel}, for readers not familiar with bifurcation theory. Note that Fig \ref{Fig:ScehmaModel} is only qualitative. In contrast, bifurcation analysis gives us more insights on bursting mechanism, where actually two regimes are possible, and quantitative results, especially on the parameters determining the transitions inducing bursting and the range of these parameters. To our best knowledge this is the first model able to provide a mechanism to account for the repetitive firing of SACs, which is the condition under which calcium controls the after-hyperpolarization phase.}

\subsection*{Explaining the wide range of interburst intervals (IBI) across species by a unique mechanism}\label{sec:Inh_IB} 
\begin{figure}[!htpb]
\centering
\includegraphics[width=18cm,height=9cm]{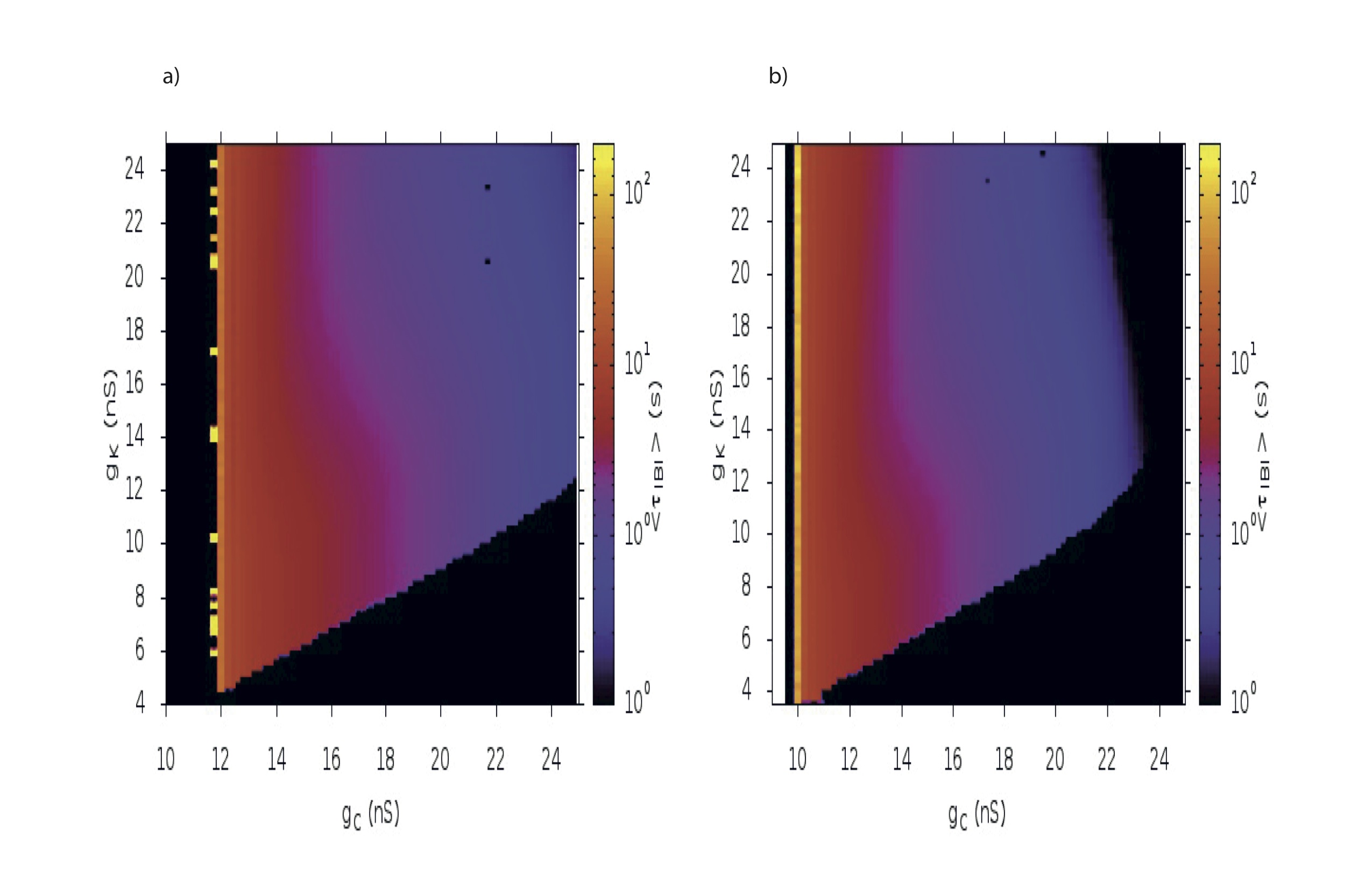} 
\caption{\label{Fig:HeatMapgCagK} \textbf{Heat map of the interburst period as a function of $g_C$ and $g_K$.}  We sample  $g_C, g_K$ on a grid with resolution $0.25$ nS. For each point, we generate $20$ trajectories of duration $2000$ s and count the number of bursts in each trajectory. A period of high calcium activity has to last at least $1$ second to be considered as a burst. The heat map show the average value of $\tau_{IBI}$ in color log scale for a fixed amount of noise $\sigma=4$ pA $ms^{1/2}$. \textbf{a)} $I_{ext}=-4$ pA (noise driven bursting) ; \textbf{b)} $I_{ext}=0$ pA (dynamically driven bursting). }
\end{figure}

Spontaneous bursting activity in immature SACs has been consistently observed across various species i.e. mice, rabbits, chicks, turtles, macaques etc., although, the measured mean interburst interval ($\tau_{IBI}$) strongly varies \cite{sernagor-grzywacz:99,maccione-hennig-etal:14,feller-butts-etal:97,sernagor-eglen-etal:00}. \bb{This experimental observation raises the question whether there exist several bursting mechanisms changing from species, explaining these variations in interburst. Bifurcations theory provides another explanation though. As we showed, in our model, bursting involves a saddle-node bifurcation. It is also known from \cite{iwata-shinichi:10}, that interburst is known to be sensitive to parameters variations near this type of bifurcation. In Fig \ref{Fig:HeatMapgCagK}, we illustrate this effect by numerically computing the heat map of the mean interburst intervals of SACs, while varying the parameters $g_K$ and $g_C$, for both bursting scenarios (noise and dynamic driven). We show that the IBI increases when approaching the Saddle-Node line. Note that we define a burst numerically by a thresholding  on $Ca^{2+}$: a burst occurs when this concentration exceeds $150$ nM during a time larger than $1$ s. There is a "critical" region (yellow zone) where interburst increases (up to several minutes). This is nicely explained by a bifurcation analysis (see section Robustness with respect to parameters variations). The irregular shape of this region -  alternation between yellow and black regions - is due to our numerical procedure. We generate $20$ trajectories of duration $2000$ s and count the number of bursts in each trajectory. In our simulation, when going close to the critical region we observe that this sampling is not sufficient; we have rare bursts giving poor statistics, and, in some cases (black regions in-between yellow ones), we observe no burst in the sample.   }

\begin{figure}[htbp!]
\includegraphics[width=\linewidth]{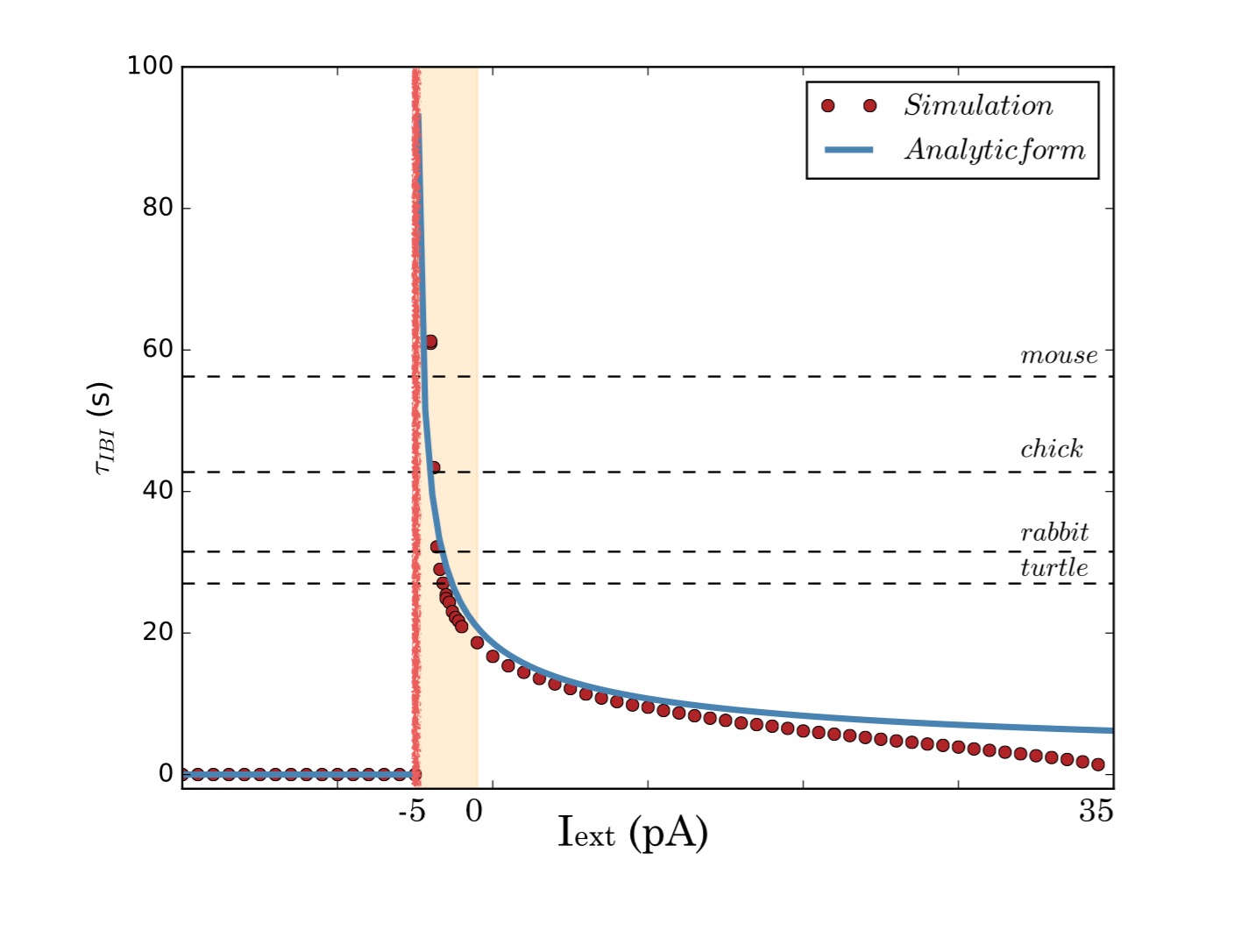}
\caption{\label{Fig:tausVL} \textbf{The behavior of the bursting period near a saddle-node bifurcation could explain interburst variability.} \textit{Red:} Computing the dependence of $\tau_{IBI}$ upon the variation of $I_{ext}$ with our model. \textit{Blue:} A 2 parameters fit of the simulated data with the curve $\frac{K}{2\sqrt{I_{ext}-I_c}}$ for $I_{ext} \in [-20, 40]$ pA. Note that this fit does not hold true away from the asymptote. (see Methods).  \textit{Red:} Values of experimentally measured interburst intervals for different species mapped to our results (see text).}
\end{figure}

To illustrate the sensitivity of $\tau_{IBI}$ to a parameter variation, we now vary the external current $I_{ext}$ (Fig \ref{Fig:tausVL}). This indeed induces strong variations in $\tau_{IBI}$, compatible with the variations observed across species, which increases monotonously and sharply as $I_{ext}$ decreases, following a hyperbola whose form has been derived analytically (see Methods and Fig \ref{Fig:tausVL}, blue trace). As shown in the paper \cite{iwata-shinichi:10}, hyperbola is one of the 2 possibilities, corresponding to the case where the noise amplitude is small compared to the excitability. We also observe a sharp transition from a bursting to a non-bursting regime, where  $\tau_{IBI}=0$, which corresponds to the loss of SACs excitability.

Based on the shape of our theoretical curve, we show that  $\tau_{IBI}$ exhibits a \textit{strong asymptotic behavior} around a very narrow regime of $I_{ext}= \in [-5, 0]$ pA. The analytic form of the function $\tau_{IBI}=f(I_{ext})$ is the following (see Methods for detailed derivation):
\begin{equation} \label{eq:tauIBI}
\tau_{IBI} = \left\{
\begin{array}{ll}
  0, & I_{ext} \leq I_c; \\
     \frac{K}{\sqrt{I_{ext}-I_c}}, & I_{ext} > I_c; \\
\end{array}
\right. 
\end{equation}
(where $K = 0.657$  s $pA^{\frac{1}{2}}$ 
and  $I_c= -5$ pA, 
for $\sigma=4$ $pA$ $ms^{1/2}$, $g_K=10$ nS, $g_C=12$ nS). The values $K, I_c$ are obtained from fit. Note that for $I_{ext} \leq I_c$, $\tau_{IBI}$ is not defined because neurons do not burst anymore. \\

This result suggests that, depending on slight variations of physiological conditions, a SAC could exhibit very variable bursting periods, providing an explanation how different species exhibit variability in the interburst interval of immature SACs.
In real SACs, the inhibitory current $I_{ext}$ could correspond to a change in the intrinsic inhibitory ionic currents of individual SAC or to external inhibitory inputs from other layers of the retina. It could also mimic changes in synaptic inhibition (e.g. $GABA_A$ maturation, changes in $[Cl]_{in}$). These  factors could drive the disappearance of bursting activity and subsequently the loss of SACs excitability.
On the other hand, an increase of $I_{ext}$ leads to a gradual decrease of $\tau_{IBI}$ towards zero, meaning that in this regime of parameters, SACs tend to burst repeatedly, without a refractory period. This scenario is not observed under normal physiological conditions, but could be tested experimentally by varying $I_{ext}$ as an externally applied current.
Finally, in the experimental paper \cite{zhou:98}, it is  found that the bursting period increases upon maturation (experiment in P1-P2 rabbit SAC). In light of the present model-driven analysis, this could be linked to a change of intrinsic properties of SACs during development.
 
To compare with experimental results, we show in the same figure $\tau_{IBI}$ for different species as found in the literature (Fig \ref{Fig:tausVL}). We have not been able to find $\tau_{IBI}$ for turtles and chicks. We found the mean inter-wave intervals $\tau_{IWI}$ instead. In order to extrapolate to $\tau_{IBI}$ in these cases, we used a common constant scaling factor of $3$, based on the ratio between $\tau_{IBI}$ and $\tau_{IWI}$ for rabbits and mice found in \cite{zheng-lee-etal:06} and \cite{ford-feller:12} respectively. Besides this narrow regime a decrease of $I_{ext}$ results in a sharp disappearance of bursting activity, which means that bursting activity stops after a limit value.



To conclude, our study suggests that bursting activity in immature SACs could share a common mechanism across species. The same mechanism could also induce interburst variability within a network SACs, thus with cells bursting with a different period. Such variability has been implemented in \cite{ford-feller:12} by varying the decay time of the sAHP from a Gaussian distribution. In our case, the variability arises without adding an extra mechanism and is induced by the hyperbolic behavior observed near a saddle-node bifurcation upon a tiny variation of a single parameter, $I_{ext}$, interpreted as an effective depolarizing/hyperpolarizing current. \bb{Taken together, the vast variability of SACs bursting features accross different species, could be explained simply by our hypothesis that the rest state of SACs lie within a narrow regime around a saddle-node bifurcation point.}

\begin{figure}[!ht]
\centering
\includegraphics[width=18cm,height=10cm]{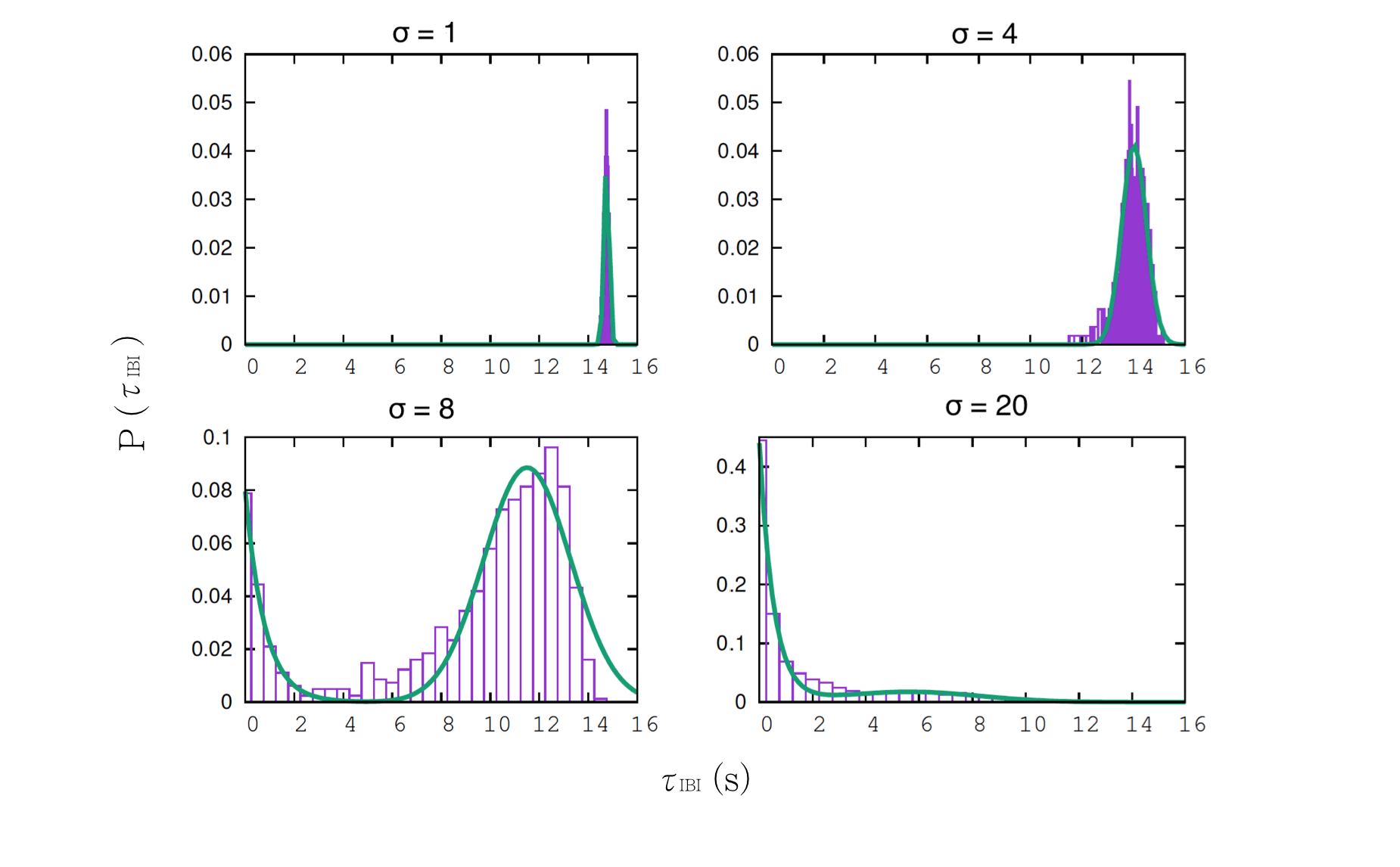}
\caption{\label{Fig:Bursting_Case2} \textbf{Histogram of bursting periods for different noise levels.} We show the distribution of interburst interval distribution $\tau_{IBI}$ with different levels of noise. \textit{Top left:} $\sigma=1$ $pA$ $ms^{1/2}$. \textit{Top right:} $\sigma=4$ $pA$ $ms^{1/2}$. \textit{Bottom left:} $\sigma=8$ $pA$ $ms^{1/2}$. \textit{Bottom right:} $\sigma=20$ $pA$ $ms^{1/2}$. Green curves correspond to fit either by a Gaussian (top), or a linear combination of a Gaussian and a decaying exponential (bottom). }
\end{figure}

\paragraph*{Characterizing the effect of noise on the probability distribution of bursts.} 
In \cite{zheng-lee-etal:06}, recordings show that the bursting periods of SACs are not regular and have a certain probability distribution. We obtain this effect by adding the noise $\xi$ to the dynamics of the voltage $V$ (see Eq \eqref{eq:Voltage1}). Note that noise is always present in real neurons due e.g. to fluctuations in ionic currents from the random opening of ionic channels. What we want to illustrate here is the role played by the intensity of this noise, controlled by the parameter $\sigma$. For small $\sigma$, dynamics has fluctuations around the deterministic trajectory with little effects during the bursting phase. In contrast, low additive noise during the slow (after-hyperpolarization) dynamics, is enough to accelerate the start of a burst. For higher values of $\sigma$ the bursting period of cells decreases dramatically. As shown in Fig \ref{Fig:Bursting_Case2} the presence of noise has also a drastic impact on the shape of the interbursts intervals distribution. 


Experimentally, maybe due to the lack of sufficiently large samples in \cite{zheng-lee-etal:06}, the exact shape of the distribution of the interburst intervals is not sharply defined.
Therefore, the comparison with our theoretical results is difficult. However, the experimental distribution obtained by these authors is definitely not exponential. As we observe an exponential distribution for large enough $\sigma$ values, Fig \ref{Fig:Bursting_Case2}, this remark provides us with an upper bound on the level of noise which should be no greater than $8$ $pA$ $ms^{1/2}$. Note that the left-wise peak in the bimodal distribution ($\sigma \sim 8$ $pA$ $ms^{1/2}$), corresponds to short bursts. Similar bursts have been observed in the spontaneous activity of retinal ganglion cells during waves, in turtle and in mouse, see e.g. control in Fig 3 and 7 in \cite{sernagor-grzywacz:99} (E. Sernagor private communication).

\subsection*{The role of the fast potassium conductance in bursting activity }\label{sec:roleK}
We now address in further detail the potential role of the fast voltage-gated potassium channels used in our model to produce bursting (see also the work of E. Marder and collaborators, in a different context \cite{marder-goillard:06}). Zheng et al. \cite{zheng-lee-etal:06} propose that the ionic channels mainly involved in the bursting activity of SACs during early development are voltage-gated $Ca^{2+}$ channels. In this work the hyperpolarizing current involved is not characterized and these authors don't mention potassium channels before the end where they perform experiments with TEA (see below for more details).
We  proposed above fast voltage-gated $K^+$ channels as a source of fast inhibition necessary for the active phase of bursting of SACs. We now justify this claim based on several experiments made by Zheng et al., interpreted in the context of our bifurcation analysis. 

\begin{figure}[!ht]
\centerline{
\includegraphics[width=0.6\linewidth]{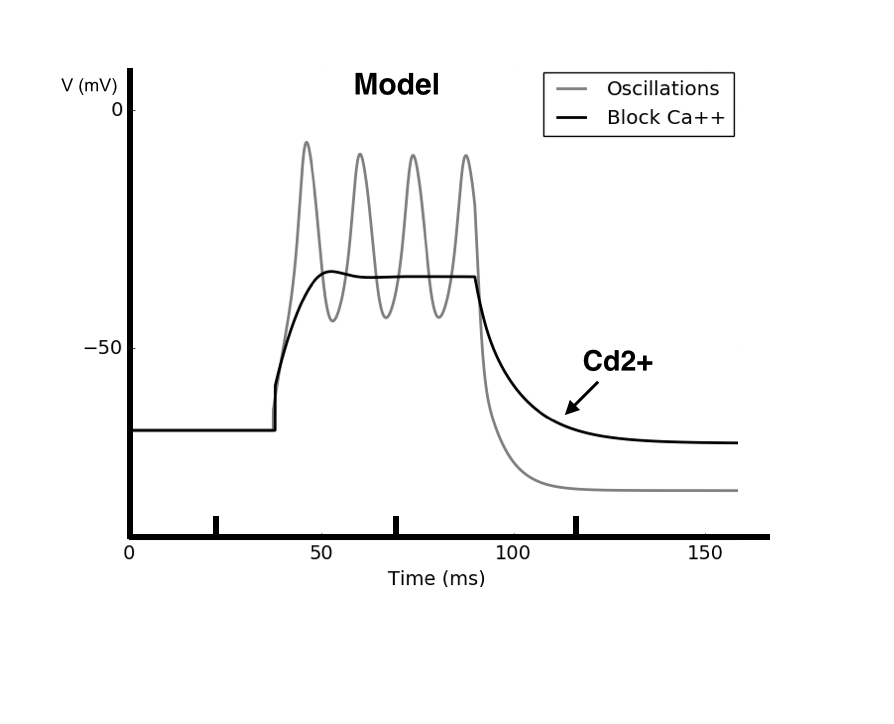}
}
\caption{\label{fig:Zhou-Kmodel} \textbf{Modeling the cellular mechanisms generating bursting activity in immature SACs.} We reproduce the experimental investigation of the cellular mechanisms involved in bursting activity made by \cite{zheng-lee-etal:06}. \textit{Gray:} Fast sub-threshold oscillations and subsequent AHP generated by a short pulse of current of amplitude $150$ $pA$ and duration $60$ ms. \textit{Black:} Blocking of all $Ca^{2+}$ related channels by $Cd^{2+}$. No oscillations are exhibited.  
}
\end{figure}

We first show (Fig \ref{fig:Zhou-Kmodel}) how our model accurately reproduces a key experiment of \cite{zheng-lee-etal:06}. Here, the authors artificially control the triggering of fast oscillations by applying a short current pulse ($I_{ext}=150$ $pA$ for $60$ $ms$) to individual immature SACs. Also, upon the pharmacological application of $Cd^{2+}$, which blocks all $Ca^{2+}$ related channels (voltage-gated $Ca^{2+}$ and sAHP), they show that no oscillatory activity is triggered upon stimulation, but only a raise in the plateau of the level of the voltage. 
The corresponding figure of their experiment in the paper \cite{zheng-lee-etal:06} has been reproduced in Fig \ref{fig:Zhou-Kmodel}A (with the kind authorization of the authors). As shown in Fig \ref{fig:Zhou-Kmodel} B, we are able to reproduce reliably this result using the fast $K^+$ channels described in Eq \eqref{eq:IK}. Particularly, we are able to simulate the emergence of fast oscillations during a short current pulse, with an AHP phase after the end of the pulse (see gray curve in Fig \ref{fig:Zhou-Kmodel} B). 
We emulate as well the disappearance of the oscillations observed by these authors, upon blocking all $Ca^{2+}$ related channels (voltage-gated $Ca^{2+}$ and sAHP), by setting, in our model, the corresponding conductances to zero (see black curve in Fig \ref{fig:Zhou-Kmodel} B).

\begin{figure}[!ht]
\centerline{
\includegraphics[width=18cm,height=16cm]{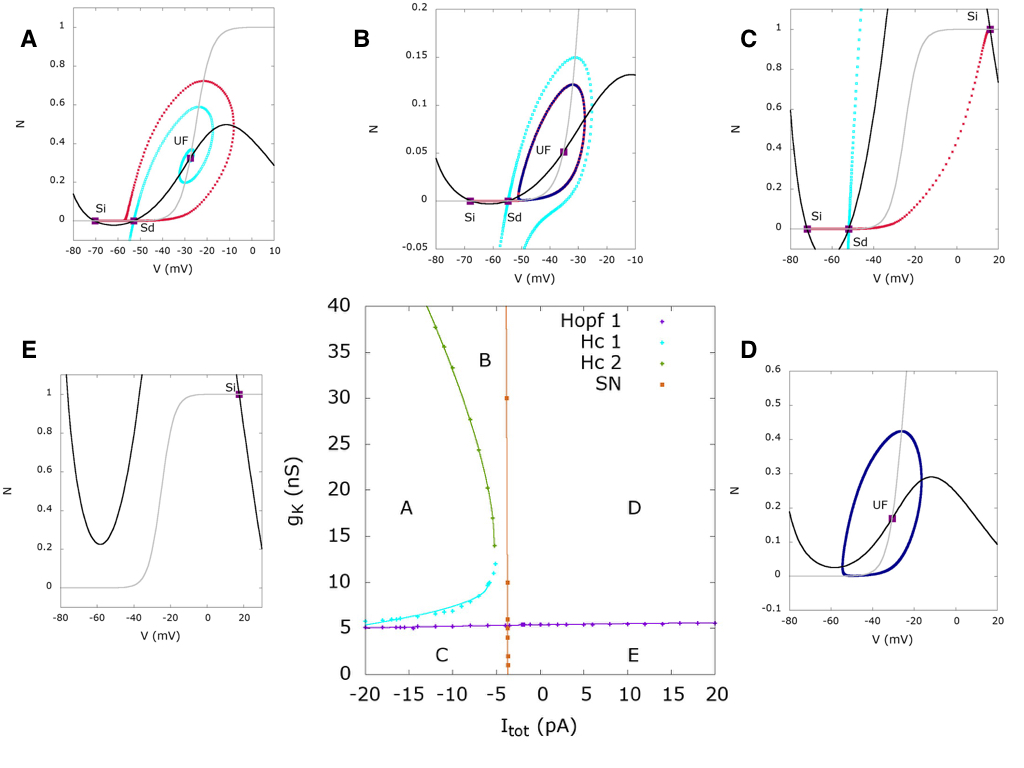}
}
 \caption{\label{Fig:BifurcationDiagram_Iext_gK_VL-70} \textbf{Bifurcation diagram as a function of $I_{tot}$, $g_K$.} At the bottom center, the bifurcation map is plotted. Points are sampling points obtained by a direct inspection of the phase portrait whereas continuous lines correspond to fit. "Hc" means "homoclinic bifurcation" and "SN" stands for "saddle-node". The capital letters A, B, C, D, E inside correspond to the typical phase portraits surrounding the bifurcation map. In these phase portraits, continuous lines correspond to nullclines ($N_V$, in black, is the  $V$-nullcline whereas $N_N$, in gray, is the $N$-nullcline). 
 'Si' stands for 'Sink', 'Sd' for 'Saddle', 'SF' for 'Stable Focus' and 'UF' for 'Unstable Focus'. Sink and stable focus correspond to stable rest state: a small perturbation about this state decay exponentially fast. On the opposite, saddle and unstable focus are unstable. For saddles we show the stable and unstable directions (black lines) as well as the stable (cyan) and unstable (red) manifolds $W_s, W_u$. 'SPo' means 'Stable Periodic Orbit'. It corresponds to fast oscillations, plotted in dark blue. In region A, there is a low voltage stable state. In region B, a stable state with low voltage coexists with a limit cycle (fast oscillations) separated by an unstable state. When the cell is in the low voltage state, a large enough perturbation (e.g. noise or other cells action) leads it to fast oscillations. In region C, two stable states, one with low voltage and one with high voltage, coexist separated by an unstable point. In region D, the cell only exhibits a fast oscillations regime, reached whatever the initial condition.  Finally, in region E, there is a stable state with high voltage. When varying parameters, continuous change occurs in the interior of a region (e.g. the period of oscillations continuously varies), whereas crossing the bifurcation lines leads to abrupt changes.}
\end{figure}

This is interpreted as follows from the bifurcation diagram Fig \ref{Fig:BifurcationDiagram_Iext_gK_VL-70},  in the plane $I_{tot},g_K$. The simulation has been done in the noise driven bursting regime where the cell, initially in its rest state, is in region A. A pulse of current $I_{ext}=150$ pA drives the cell in region D where it fast spikes. Removing the current pulse drives back the cell in region A. Here we observe a small sAHP hyperpolarizing the cell. 
In order to observe the same behavior in the dynamically driven bursting regime, one needs to clamp the voltage so as to maintain the cell in the rest state.


These observations support therefore our proposition that the hyperpolarizing component of the fast oscillations observed in immature SACs are driven by fast voltage-gated $K^+$ channels.

\subsection*{Exploring the role of the potassium conductance in the loss of SACs excitability upon maturation} \label{sec:rolegK_maturation} 

SACs in the retina of vertebrates lose their ability to spontaneously burst once they reach a certain stage of development different for each species. This transient excitability is a key process in the developing retina and the shaping of the visual system. It is not yet clear which physiological properties of SACs change upon maturation causing the abrupt change from autonomous bursting to rest state. So far, to our best knowledge, no experiment has studied in detail the biophysical properties of immature SACs at the level of each ionic channels involved during this window of development. However, there are some indirect experimental implications which help us extrapolate a possible scenario on how SACs change their properties upon maturation. In \cite{ozaita-petit-jacques-etal:04} Ozaita et al. show that mature SACs not only stop bursting but also cannot be depolarized beyond $-20$ mV. A specific type of inhibitory voltage gated $K^+$ channels, $Kv_3$, is responsible for this property, providing an electric shunt to SACs somas. In the contrary, in the developing retina, SACs are autonomous bursters, depolarised beyond mature SACs ($\sim -5$ mV) \cite{zheng-lee-etal:06}. Therefore, upon maturation, the characteristics of voltage gated $K^+$ channels evolve, leading to a drastic change in SACs activity. In our model, these observations can be reproduced by the variation of two parameters; i) the conductance $g_K$  and ii) the half-activation potential $V_3$ of the voltage gated $K^+$ channels. 

Here a point of caution is necessary. The modeling chosen here, in the same spirit as Morris-Lecar or Hodgkin-Huxley model, characterizes the behavior of a small piece of a cellular membrane, large enough to contain many ionic channels so that the description in terms of activity variables ($N,R,S$) is correct. As a consequence, parameters such as $g_K, V_3$ are phenomenological, depending on membrane characteristics (e.g. the density of ionic channels of a given type), and on channels characteristics. When we vary a parameter such as $V_3$, it means that we  mathematically change the threshold of the term mimicking fast-$K$ channels in the model. The biophysical interpretation of this variation is more difficult. The easiest interpretation is that it corresponds to a change of threshold response of the corresponding channels, but other explanations could be also possible. Assume for example that there are two types of fast $K^+$ channels (e.g. kV3.1 and kV3.2) on the membrane, with different densities and reactivity. Then, a mere change of their density during development would impact the value of the phenomenological parameters $g_K, V_3$ of our model. \\

\begin{figure}[!httbp!]
\includegraphics[width=\linewidth]{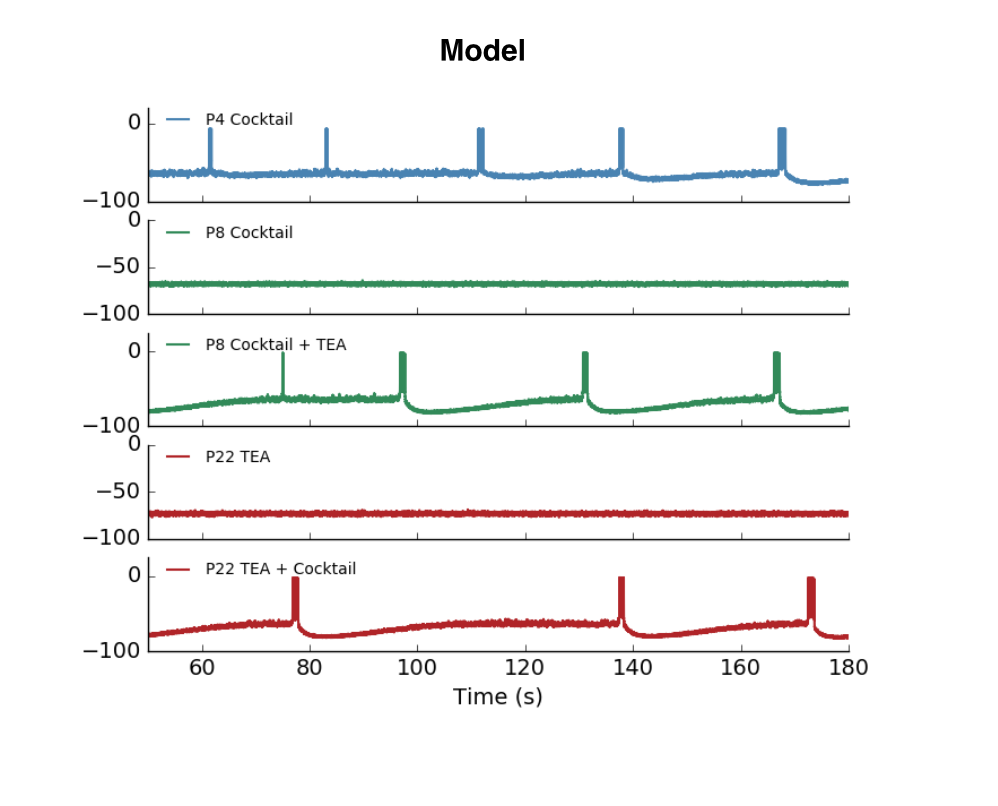}
\caption{\label{Fig:P4}
 \textbf{Varying the potassium conductance $g_K$ and the half-activation potential $V_3$ changes the excitability of SACs in our model as in the pharmacological manipulations made by \cite{zheng-lee-etal:06}.} \textit{Blue:} Modeling the bursting activity of isolated P4 SAC, $gK=8$ nS, $V_3=-16$ mV. \textit{Green top:} Modeling P8 isolated SACs where we see no bursting activity, $gK=10 nS, V_3=-34 mV$. \textit{Green bottom:} Decreasing $gK=8$ nS, $V_3=-34$ mV we restore oscillations. \textit{Red top:} Modeling P22 coupled SAC where there is no bursting activity upon treatment with TEA. An inhibitory current $I_{ext}=-10$ pA is applied to coarsely mimic the inhibitory input to SAC at this stage of development. To account for the TEA application we take decrease the conductance $gK=4.5$ nS ($V_3=-35$ mV ). \textit{Red bottom:} Restoration of bursting at P22 by removing all inhibitory synaptic connections, $I_{ext}=0$ pA. Removing inhibition from mature amacrine circuitry, along with blocking a subfamily of $K^+$ channels with TEA, is enough to re-initiate bursting activity. The values proposed here for $V_3, g_K$ are only indicative, as the same behavior is observed in a wide range of parameters (see the bifurcation diagram in Fig  \ref{fig:HeatMapIBI_vs_gK_V3_theta1_sigma4_VL-72}).}
\end{figure}

\begin{figure}[!ht]
\centerline{
\includegraphics[width=18cm,height=10cm]{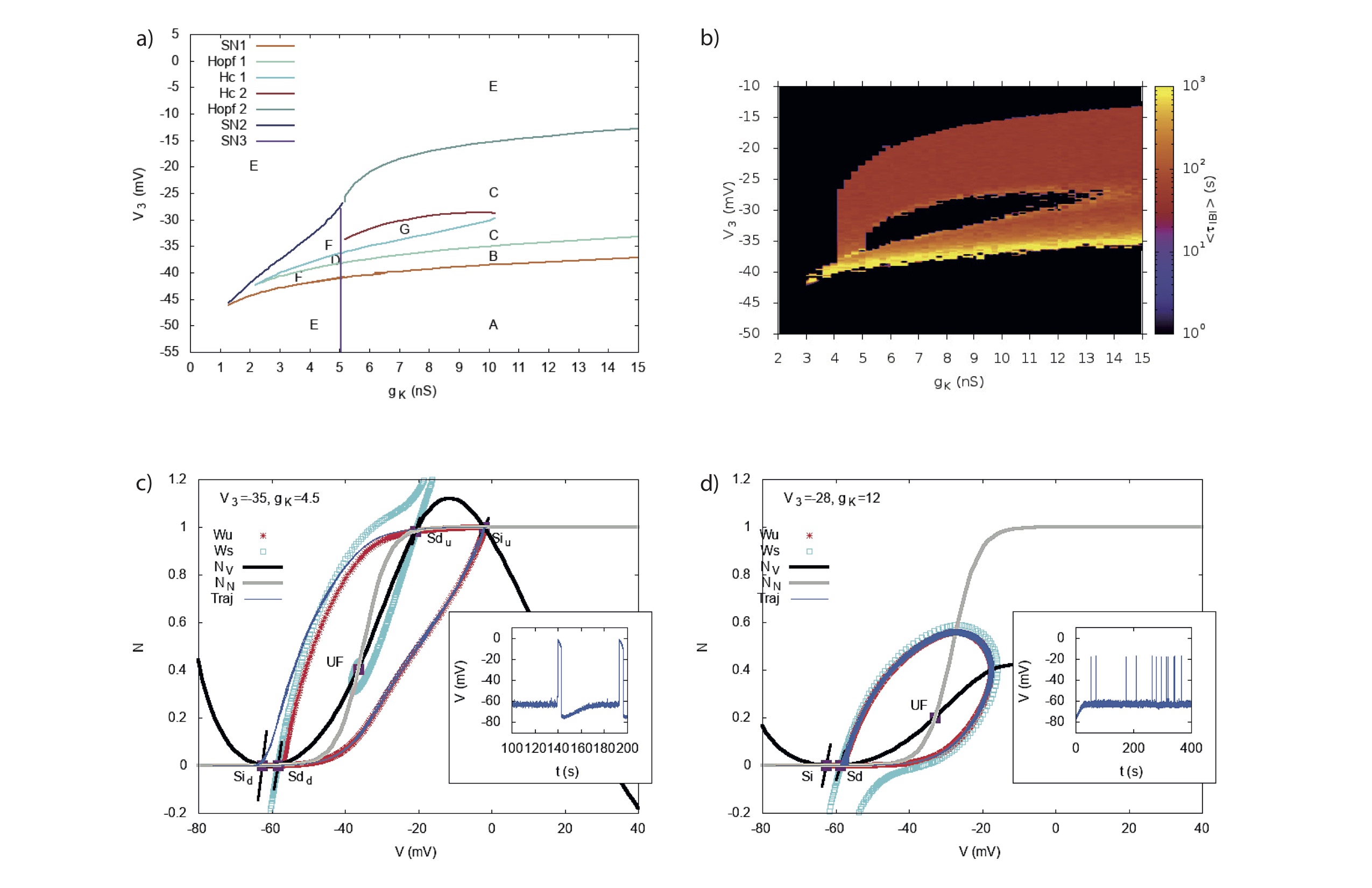}}
\caption{ \label{fig:HeatMapIBI_vs_gK_V3_theta1_sigma4_VL-72} \textbf{Bifurcation diagram and heat map of $\tau_{IBI}$ in the plane $V_3 - g_K$. a)} In region A, there is a stable fixed point  with low voltage. In B, 2 stable fixed points coexist with a saddle-point. In C a stable fixed point with low voltage coincides with a stable periodic orbit giving rise to fast oscillations. In region D  there are 4 fixed points (2 stable, 2 unstable) and a limit cycle. In region E there is a stable fixed point with low voltage and a stable fixed point whose position depends on $g_K$ and $V_3$. In region F there are five fixed points, 2 stables. Region G contains 2 stable fixed point (a sink and a focus) separated by an unstable fixed point. Bursting takes place in regions C and D.
\textbf{b)} Heat map of the interburst interval $\tau_{IBI}$. Same representation as Fig \ref{Fig:HeatMapgCagK}. We sampled the plane $g_K,V_3$ with a step of $0.25$ nS on the $g_K$ axis and $1$ mV on the $V_3$ axis, explaining the irregular shape of the  border.  \textbf{ c) and d)} There is a discrepancy between the heat map and the bifurcation diagram: (i)
The bifurcation map does not predict bursting for $g_K \in [4,5]$ nS, $V_3 \in \sim [-40,-25]$ mV whereas the heat map shows  calcium peaks lasting more than 1 s in this region. This is due to the particular structure of the phase portrait of the Morris-Lecar model in this region and the effect of noise, as shown in \textbf{c)}. This phase portrait has two saddles ($Sd$) down ($Sd_d$) and up ($Sd_u$), and two sinks ($Si$), down ($Si_d$) and up ($Si_u$). The unstable manifold ($W_u$, in red) of $Sd_d$ connects to $Si_u$, on the right, and to $Si_d$, on the left; The unstable manifold of $Sd_u$ connects to $Si_u$, on the right, and to $Si_d$, on the left. Thanks to noise, dynamics can leave $Si_d$ and reach $Si_u$ where it stays until sAHP rising drive it back to $Si_u$. This is illustrated by the blue trajectory (complete system with noise) which follows the unstable manifolds of the Morris Lecar model. The inset shows the trajectory of the voltage for the complete system in the presence of noise. (ii) \textbf{d)} While the bifurcation diagram predicts bursting for $V_3 \sim -28$ mV, $g_K \sim 12$ nS, the heat map does not show bursting in this region (black). This is again due to noise. When the cell is on low rest state (Si) noise drives it to the limit cycle, but after a few oscillations, noise drive the cell back to the rest state Sd (see inset for the voltage trajectory). This effect depends on the distance between the two fixed points Si, Sd, fixed by the model parameters, and on the noise intensity $\sigma$ (here $4$ pA ms$^\frac{1}{2}$).}
\end{figure}
With our model, we are able to reproduce an experiment by Zheng et al. 2006 \cite{zheng-lee-etal:06}. The results of their experiment are shown in Fig \ref{Fig:P4}, left column (with the kind authorization of the authors), whereas the model results are presented in Fig \ref{Fig:P4}, right column. Our results are based on a bifurcation analysis  in the plane $V_3 - g_K$, which reveals a wide region of parameters where bursting takes place.  In Fig \ref{fig:HeatMapIBI_vs_gK_V3_theta1_sigma4_VL-72} we show, on the top left, the bifurcation diagram in the plane $V_3, g_K$ and on the top right, a heat map, similar to  Fig \ref{Fig:HeatMapgCagK}). This map makes easier the interpretation of Zheng et al. experiment which mainly addresses the following questions:
\begin{itemize}
\item \textit{How does SACs autonomous bursting stop?} Zheng et al. consider first isolated SACs (all synaptic connections are inactivated with a pharmacological cocktail). While SACs are bursting spontaneously (i.e. without the influence of the other cells) at P4 (first row, left in Fig \ref{Fig:P4} ), this spontaneous activity disappears at P6 (second row, left). 
 We can easily reproduce this observation (Fig \ref{Fig:P4} , first and second row, right) by moving from region C (fast oscillations) in the bifurcation diagram of Fig \ref{fig:HeatMapIBI_vs_gK_V3_theta1_sigma4_VL-72} to another region. For example, a simple transition to region A is obtained by decreasing the half-activation potential $V_3$ of the fast $K^+$ channels, leading to the saturation of SACs depolarization below $-20 mV$ observed in mature SACs. A concomitant variation of $g_K$ is also possible leading to stop bursting as well. To illustrate this scenario, we set $g_K=10$ nS and $V_3=-34$ mV as an indicative example (see Fig \ref{Fig:P4}). 

\item \textit{How can bursting be restored in further mature SACs?}  Zheng et al. \cite{zheng-lee-etal:06} show that bursting activity can be re-initiated pharmacologically in P8 (stage III) and P22 retinas with chemical agents upon i) blocking all synaptic connections  (gap junctions, cholinergic, gabaergic and glutamatergic synapses), especially inhibition by applying an antagonist cocktail and ii) decreasing the conductance of the fast inhibitory $K^+$ channels, by applying tetraethylammonium (TEA). This suggests that upon maturation, bursting is in fact suppressed by alterations in the intrinsic properties of individual SACs. Particularly, in the same preparation (P8 isolated SACs), the authors apply a voltage-gated $K^+$ channel blocker tetraethylammonium (TEA), which results in restoring bursting activity in isolated mature SACs (Fig \ref{Fig:P4}, third row, left). A potential type of potassium channels involved is the TEA -sensitive $K^+$ of the $Kv_3$ family. Therefore, this experiment emphasizes once again the potential role of fast potassium channels and especially the $Kv_3$ family. We mimic the blocking of these channels in our model by decreasing the conductance $g_K$.  When $g_K$ is small enough, we reproduce the bursting restoration (Fig \ref{Fig:P4} , third row, right). This example corresponds, in Fig \ref{fig:HeatMapIBI_vs_gK_V3_theta1_sigma4_VL-72}, to a motion from $(g_K,V_3)=(10 nS,-34 mV)$ (region B) to $(g_K,V_3)=(8 nS,-34 mV)$ (region C). Note that the chosen value  $V_3=-34$ mV allows us to fix the maximal depolarization to $-20$ mV, the  value observed by Ozaita et al. \cite{ozaita-petit-jacques-etal:04} for further mature SACs. Remark also that the lower branch of the heat map Fig \ref{fig:HeatMapIBI_vs_gK_V3_theta1_sigma4_VL-72} (region between G and B in the bifurcation map) corresponds to interburst intervals of the order of $8-16$ min, so according to SACs recording not to a biophysically plausible bursting regime. 

\item \textit{How can bursting be restored in adult SACs?} Upon further maturation, Zheng et al. show that the restoration of bursting in P22 SACs (late stage III-before eye-opening) depends on the interplay between the change of the intrinsic properties of $K^+$ channels strong inhibition (TEA application) and blocking the strong inhibitory (gabaergic) input induced by other amacrine cells at this phase (Fig \ref{Fig:P4}, fourth and fifth row, left). Neither of these conditions suffice alone for the bursting restoration. To model the gabaergic inhibitory input to coupled SACs at P22, we add an inhibitory current $I_{ext}=-10$ pA whereas the effect of TEA is modeled by decreasing the $g_K$ conductance to $4.5$ nS. This conductance is not set to $0$ because TEA blocks only one subtype of $Kv_3$ channels while there exist several of them (see below) in the developing retina. The whole operation has the effect of suppressing bursting (red trace top) and corresponds moving from region B to region A in Fig \ref{Fig:BifurcationDiagram_Iext_gK_VL-70}. Bursting re-initiation is obtained by setting back the current $I_{ext}$ to zero.

\end{itemize}
Along these lines, we have been able to reproduce the effects of the pharmacological manipulation performed by Zheng et al. at the three separate ages; $P4$, $P8$ and $P22$ in Fig \ref{Fig:P4}. \\
 
\bb{Taken together, during development, SACs exhibit a transient bursting activity which is lost upon maturation. In the context of our bifurcation analysis, this process is interpreted as follows: immature SACs rest state lies close to a bifurcation point in order to facilitate the bursting activity which is essential for the generation of retinal waves and the shaping of the visual system. Upon maturation and once this functionality is no further useful for the retinal circuitry, we predict that SACs rest state is naturally driven towards the stable branch (further left in Fig \ref{Fig:BigDiagML}A), away from the bifurcation point, where bursting generation is no longer possible (we remind that noise could drive the cell to cross the bifurcation but, the further away, the less biophysically plausible this scenario is regarding the noise amplitude). In our model, this essentially corresponds to adding a hyperpolarizing (negative) current as an input. For identifying the source of such current, we propose a conjecture on the possible role of the $Kv_3$ potassium channels during development, based on our analysis, which could reveal a mechanism of how SACs lose their excitability.}

\subsection*{A conjecture on the role of the $Kv_3$ channels in the loss of SACs excitability upon maturation}
We predict that bursting occurs in immature SACs, during a certain age window while fast potassium channels are under-expressed. Upon development, we make the hypothesis that potassium channels increase their expression and spontaneous bursting stops when the hyperpolarizing $K^+$ dominate the dynamics over the depolarizing $Ca^{2+}$ channels. Also, we suggest that upon maturation, the half activation potential $V_3$ decreases in order to provide a saturation in mature SACs activity as shown in \cite{ozaita-petit-jacques-etal:04}. This theoretical prediction could be experimentally tested, by following the expression of the fast $K^+$ channels in SACs during development and showing a potential increase of their activity along maturation (see discussion section). 
Moreover, we propose that the fast voltage gated $K^{+}$ channels involved is the $Kv_3 $ family. More specifically, two subtypes of fast voltage gated $K^{+}$ channels have been identified \cite{kaneda-ito-etal:07, ozaita-petit-jacques-etal:04}, both belonging to the $Kv_3$ family: i) $I_K$ delayed rectifier currents, sensitive to TEA (tetraethylammonium) emitted by subunits $Kv_3.1$ and $Kv_3.2$ and ii) $I_A$ $A$-type currents, sensitive to 4-amino- pyridine (4AP), emitted by subunits $Kv_3.3$ and $Kv_3.4$ \cite{kaneda-ito-etal:07} which are not sensitive to TEA. The apparition and evolution of such channels in SACs during development has not been studied yet. However, in the experiment of \cite{zheng-lee-etal:06}, the application of TEA in \textit{immature} SACs (rabbit $\sim P8$) was crucial to the re-initiation of bursting activity. This finding implies that, during development TEA sensitive $K^+$ channels are already expressed in SACs. On this basis, we propose that the specific type of $K^+$ channels responsible for the spontaneous bursting activity during development and eventually its loss upon maturation, is the $I_K$ rectifier currents of the $Kv_3.1$ and $Kv_3.2$ subtype.\\


 To conclude, we propose that, during early development ($<P6$), $Kv_3$  channels are under-expressed, allowing a competition between inhibition ($K^+$) and excitation ($Ca^{2+}$), leading SACs to burst (see Fig \ref{Fig:P4}, blue trace). Upon maturation, the expression of these channels could evolve increasingly, leading to stronger inhibition, dominating fully the competition of inhibitory/excitatory channels, by suppressing oscillations completely. Therefore, we suggest that the evolution of the expression of fast $Kv_3$ channels could be part of the transient process that leads to a complete loss of excitability of mature SAC. Our theoretical results indicate that the level of the expression of $Kv_3$ channels could increase gradually upon maturation, which essentially means that the conductance of these channels would increase as well as other physiological properties of the channels such as kinetics parameters like the half-activation potential (fixing the characteristic activation sigmoid of the channel). This type of experiment could elucidate the exact role of $Kv_3$ channels in the intrinsic properties of the excitability of SACs. \\

\subsection*{Robustness with respect to parameters variations.}
\bb{We now demonstrate the robustness of this analysis to variations of our main parameters.
We already showed the effect of varying $I_{tot}$ and $g_K$ in Fig. \ref{Fig:BifurcationDiagram_Iext_gK_VL-70}. As bursting is a competition between calcium and potassium dynamics, we have also checked the structure of bifurcations when varying the calcium conductance $g_{C}$ and the potassium conductance $g_K$ around the values we have fixed in our model.} We have used different software (XPPAUT \cite{ermentrout:02}, MATCONT \cite{dhooge-govaerts-etal:03}, PyCont \cite{clewley:12}) and we were not able to get all bifurcations with them (especially homoclinic bifurcations). So we obtained the bifurcation diagram by a direct inspection of the phase portraits with respect to parameters variations. The points shown in the bifurcation maps are sampling points whereas continuous lines correspond to fit. We note that Hopf and Saddle Node bifurcations were obtained identically using the bifurcation analysis software quoted above.\\

\begin{figure}[!ht]
\centerline{
\includegraphics[width=18cm,height=16cm]{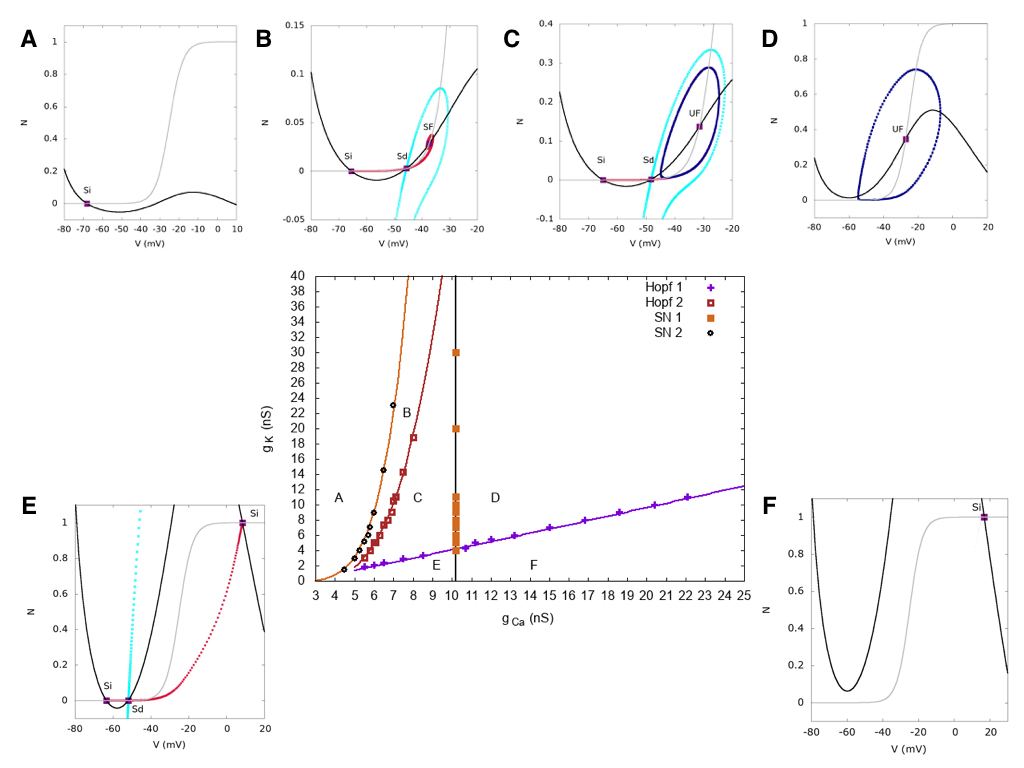}
}
 \caption{\label{Fig:BifurcationDiagram_gCa_gK_VL-70} \textbf{Bifurcation diagram as a function of $g_C$, $g_K$ in the dynamically driven bursting regime.} Same representation as Fig \ref{Fig:BifurcationDiagram_Iext_gK_VL-70} and same comments. Note the correspondence with the heat map Fig \ref{Fig:HeatMapgCagK}.
}
\end{figure}

\begin{figure}[!ht]
\centerline{
\includegraphics[width=18cm,height=16cm]{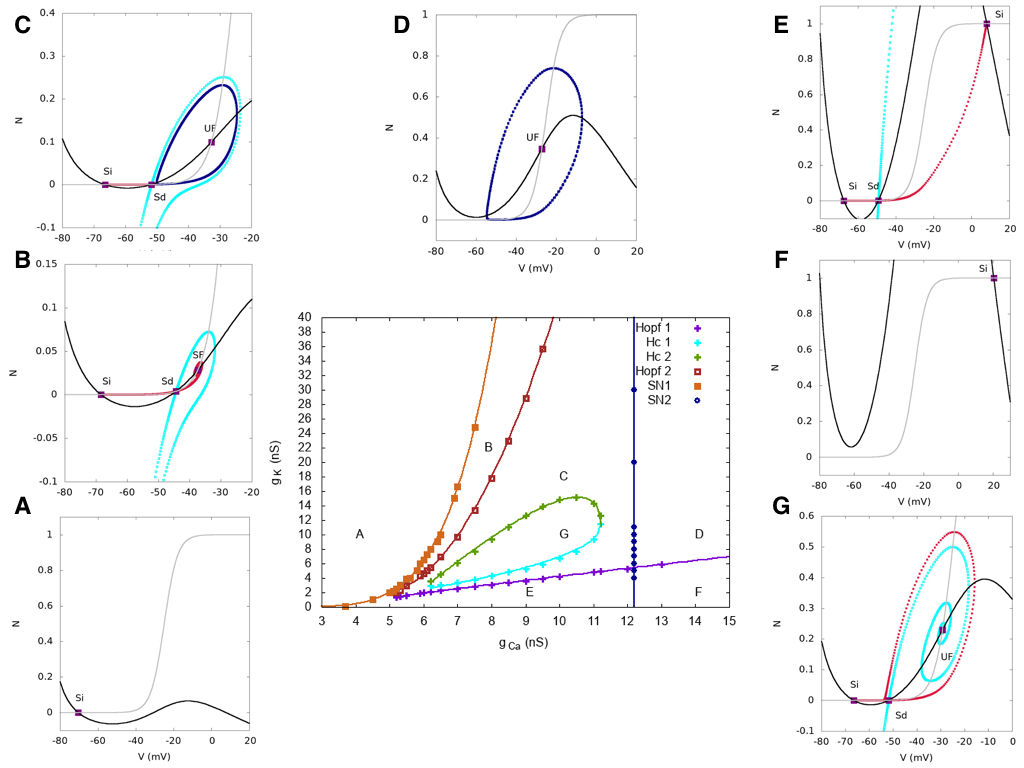}
}
 \caption{\label{Fig:BifurcationDiagram_gCa_gK_VL-72} \textbf{Bifurcation diagram as a function of $g_C$, $g_K$ in the noise driven bursting regime.}  Same representation as Fig \ref{Fig:BifurcationDiagram_Iext_gK_VL-70}. In region A, there is a unique stable rest states with low voltage. In region B two stable rest state with low voltage coexist separated by an unstable point; hence dynamics is bistable. In region C, a stable rest state with low voltage coexists with a limit cycle (fast oscillations) separated by an unstable state. When the cell is in the low voltage rest state, a large enough perturbation leads it to fast oscillations. In region D the cell only exhibits a fast oscillations regime, reached whatever the initial condition. In region E two stable rest states, one with low voltage and one with high voltage, coexist separated by an unstable point. Region F has only one high voltage stable rest state. Finally, in region G, there is a low voltage rest stable state. }
\end{figure}

\bb{The resulting bifurcation diagrams, in the two regime of bursting, are drawn in Fig \ref{Fig:BifurcationDiagram_gCa_gK_VL-70} (dynamically driven bursting) and \ref{Fig:BifurcationDiagram_gCa_gK_VL-72} (noise driven bursting). In the noise driven bursting regime there is a region (G in the Fig \ref{Fig:BifurcationDiagram_gCa_gK_VL-72}), delimited by two homoclinic bifurcations lines, where the cell is bistable without fast oscillations. Note that this region does not exist in the dynamically driven bursting regime, hence Fig \ref{Fig:BifurcationDiagram_gCa_gK_VL-70} and Fig \ref{Fig:BifurcationDiagram_gCa_gK_VL-72} have a different $x$ axis range. (The goal is to enlarge region G, which does not appear in Fig 6.)}\\

\section*{Discussion}

In this paper we have proposed a model for spontaneous bursting of SACs, one of the key ingredients for the initiation of stage II retinal waves. To our best knowledge, this is the first model able to characterize SACs dynamics in great detail and to show that biophysical properties of SACs are sufficient to explain intrinsic bursting and refractoriness in these cells. The model can reproduce spontaneously occurring fast oscillatory activity followed by long refractory periods in individual SACs. It is explained by bifurcations arising purely from intrinsic excitable membrane properties. SACs can spontaneously switch to strong oscillatory activity followed by refractoriness, reflecting what has been reported experimentally in immature SACs during the period of retinal waves. The bursting frequency is determined by a Ca-activated K conductance responsible for the sAHP.
This led us to propose several conjectures and possible experiments, directly inspired from the model analysis. 

We would like now to develop other aspects, not considered in the main text.

\paragraph{Biophysical simplifications.}  Although our model is able to explain fairly well and reproduce several aspects in the physiology of bursting SACs, as well as to propose new experimental predictions it is yet simplified with respect to the real cellular processes in SACs, lacking potentially certain biological mechanisms with important consequences. Especially, there are several important ion channels (e.g. $I_{Na}$) \cite{zhou-fain:96} and transporters in immature SACs that are not considered in our mathematical description. These currents make changes in the intracellular ion concentrations (ex $Cl^-$ concentration \cite{zhang-pathak-etal:06}) and make changes in the reversal potentials (e.g. $V_L$) of the ion channel currents. A change in $V_L$ results in a change of the effective current $g_L V_L$ in eq. \eqref{eq:Voltage1} and therefore can be absorbed in $I_{ext}$.
Here, we justify further our approximations:
\begin{itemize}
\item\textbf{The role of sodium channels.} In \cite{zheng-lee-etal:06}, it has been shown that voltage-gated $Na^+$ channels do not contribute to the bursting activity of immature SACs, since upon the application of tetrodoxin (TTX), their activity remained unaltered. Note that the paper \cite{zhou-fain:96} (10 years anterior to \cite{zheng-lee-etal:06}) mention the role of $Na^+$ but it seems only responsible for the initial peak of voltage arising when the burst starts. This point is confirmed in \cite{zheng-lee-etal:06}, see Fig $4$ of this paper.  We have therefore decided to remove $Na^+$ channels from our modeling. 
Eliminating $Na^+$ channels allows us to consider  a more simplified model than Hodgkin-Huxley, yet a canonical model in computational neuroscience, the Morris-Lecar model.
\item\textbf{On calcium dynamics.} 
 In \cite{zheng-lee-etal:06, hennig-adams-etal:09}, the authors showed that the long refractory period of SACs is controlled by $Ca^{2+}$-gated $K^+$ channels, probably SK-like, as those observed in pyramidal cells in hippocampal neurons studied in \cite{abel-lee-etal:04}. Based on these experimental observations, we have proposed a model of sAHP dynamics. Although our modeling is widely inspired from \cite{hennig-adams-etal:09}, we get deeper in the biophysical description of Calcium dynamics, based on the seminal work by M. Graubner and collaborators \cite{graupner:03,graupner-erler-etal:05} (see Methods).  Additionally, the parameters were tuned to match the SK channels experimental characterization made by Abel et al. \cite{abel-lee-etal:04}. 
Yet, we have simplified the dynamics of NCX exchangers and we neglected the ionic pumps currents. These simplifications are detailed in Methods.
\end{itemize} 

\paragraph{The Interburst variation and the interpretation of $I_{ext}$.}
Spontaneous bursting activity in immature SAC has been observed consistently across species i.e. mice, rabbits, chicks etc., but with different characteristic bursting periods. In Fig \ref{Fig:tausVL}, we show the relationship of the average bursting period $\tau_{IBI}$ with respect to the value of the tunable current $I_{ext}$ whose effect could be translated on how an effective inhibition affects the characteristics of the bursting activity and even stops it completely. The source of such an inhibition could be linked either with intrinsic cell transient properties or inhibitory inputs from other layers of the retina upon maturation. Although it is unlikely that changes in intrinsic properties of SACs would occur over the short period of stage II, such changes could be observed across species. As the interburst intervals (measured in retinal ganglion cells) increase and then decrease (in mouse) during Stage II waves (see \cite{maccione-hennig-etal:14}) the effective inhibition current $I_{ext}$ is more likely arising from changes in synaptic inhibition (e.g. GABA A maturation or changes in the inner chloride concentration).

\paragraph{How identical cells can display variability in behaviour.} One of the key result of this study is that SACs display bursting because they are close to a bifurcation point.
As a consequence, the characteristics of bursting (IBI, amplitude) vary strongly upon tiny variations of physiological parameters such as $g_K,g_C, I_{ext}, V_3 \dots$ and noise can induce huge fluctuations in their bursting period. In a network of SACs those parameters may fluctuate. Especially, the presence of acetylcholine inducing a cholinergic current impacts the bursting period as was experimentally observed by Zheng et al. in \cite{zheng-lee-etal:04,zheng-lee-etal:06} and confirmed by our model \cite{karvouniari-gil-etal:16,karvouniari-gil-etal:17}.  Therefore, a network of identical coupled SACs is expected to display a huge variability: SACs are not identical bursters. 
This variability has been invoked and modeled in a paper by Ford and Feller\cite{ford-feller:12} where they show how it impacts the retinal waves dynamics and provide more realistic results than an homogeneous model. In their paper, inhomogeneity is imposed ad hoc (Gaussian distribution of IBIs) and does not evolve in time. In contrast we propose here that (i) inhomogeneity purely results from the biophysical mechanisms leading SACs to burst; (ii) this heterogeneity evolves in space and time, depending on the collective dynamics (e.g. acetylcholine concentration varies in space and time and induces a variability in a given cell bursting period). In particular, one could have two populations of cells: one population acting as dynamically driven bursters with a well defined bursting period, and one population acting as noise driven bursters with a large variation of the period. Those populations would act in a different way on the collective dynamics of waves (see next section).      

\subsection*{How inhomogeneous bursters can produce retinal waves ?}
To our best knowledge, this model is the first able to characterize SACs dynamics in great detail and to show that biophysical properties of SACs are sufficient to explain intrinsic bursting and refractoriness in these cells.
Yet, this work was concerned with individual neuron dynamics a crucial activity for the emergence of a propagating wave activity. A next step therefore consists of studying a network of coupled SACs. The crucial point is to understand how a network of coupled inhomogeneous bursters can achieve the third ingredient necessary to initiate retinal wave: synchrony. 
In the retina, at stage II, SACs are coupled via cholinergic synapses, whose strength evolve during development \cite{zheng-lee-etal:06}. 
Modeling the cholinergic coupling, we observed indeed that SACs are able to achieve local synchrony and generate waves for sufficiently large value of the cholinergic conductance, compatible with experiments \cite{zheng-lee-etal:06}. This is further developed in a forthcoming paper but we give here the main ideas. Local synchrony results from the cholinergic coupling via the classical mechanism of non linear synchronization between inhomogeneous oscillators, called  "Arnold' tongues" \cite{arnold:88}.
For a sufficiently large cholinergic conductance, compatible with experiments, this synchrony leads to a wave propagation. This propagation can be described, in our model, via a transport equation allowing to compute the wave characteristics (speed, size, duration) as well as the effect of sAHP on dynamics. Because of its slow time scale, quite slower than the characteristic time of the wave propagation, sAHP somewhat imprints the medium where the wave has to propagates leading to  transient spatial structures observed in  Ford et al. \cite{ford-felix-et-al:12}.  The full analysis of these aspects will be done in a forthcoming paper (see \cite{karvouniari-gil-etal:16,karvouniari-gil-etal:17,karvouniari:18}).

\paragraph{Pharmacological control.}
Although this model has many parameters, bifurcation analysis allows us to highlight several of them, $g_C$, $g_K$, $V_3$ and $I_{ext}$, controlling important aspects of dynamics, in direct links with experiments. Varying those parameters have a deep impact on SACs dynamics as shown e.g. in Fig \ref{Fig:P4}. This is possible via a pharmacological control (e.g. TEA to vary K conductance, Cd to vary Ca conductance, synaptic input changing $I_{ext}$ $\dots$). It would be interesting to confirm experimentally our bifurcation diagram with such experiments.

In addition, this study opens the possibility that mature SACs are still potential bursters. Then, by a suitable pharmacological treatment, guided by the bifurcation analysis, they could start to burst in mature retinas. Would this restore a wave activity as recently observed in mature mice retinas \cite{kolomiets-cadetti-etal:16} ? This deserves however further investigations.
Especially, SACs are not connected to each other in the adult retina (cholinergic synapses disappear at the end of stage II \cite{zheng-lee-etal:06}). In adults SACs make synapses with bipolar cells axons, with retinal ganglion cells, and with some other amacrines types (inhibitory) providing several potential coupling mechanism to propagate waves. 

\subsection*{\bb{Proposing new experiments towards the validation of our theoretical predictions}}

\paragraph{\bb{How to validate experimentally a bifurcation diagram for SACs dynamics?}} \bb{It would be very elucidating to conduct experiments in order to calibrate accurately SACs dynamical repertoire with respect to our bifurcation analysis shown in Fig \ref{Fig:BigDiagML}A. It would require to apply several consequent constant current pulses to an isolated immature SAC, within a wide range, and create an "experimental bifurcation diagram", showing the observed behaviors of the cell upon the variation of a constant current. Such an experiment, would shed light also on understanding the two possible drives for triggering SACs bursting which is a result of our modeling. Also, it would help validate our hypothesis about SACs being close to a bifurcation during development.}

\paragraph{\bb{How to validate experimentally the possible role of $K^+$ channels on the loss of SACs excitability?}}  \bb{One of the predictions that emerged from our model is that the $Kv_3$ channels, that are expressed in the SACs at adult stage \cite{ozaita-petit-jacques-etal:04}, may not be expressed at stage II. To test this hypothesis, immunolabeling experiments are needed in the mouse retina, using an antibody labeling the $Kv$ channels, both in adult and immature retina in order to compare and deduct whether the localization of $Kv$ channels is weaker or inexistent in the immature retina (P3-P10 for mice). Along with immunochemistry experiments, actual measurements of the $K^+$ currents at different stages of development (we thank one of the reviewers for this comment) would elucidate their possible transient expression during development and their role in the lose of SACs excitability. }

\section*{Methods}

\subsection*{Deriving sAHP dynamics}\label{App:sAHP_appendix}

To our best knowledge, the ionic channels types involved in sAHP for immature stage II SAC are not precisely known. However, Zheng et al. argue in \cite{zheng-lee-etal:06} that these channels could share characteristics with SK channels, thoroughly studied for pyramidal neurons by Abel et al. in \cite{abel-lee-etal:04}. On this basis we modeled SK channels dynamics. 
SK channels have four subunits associating to form a tetramer. The SK
channel gating mechanism is controlled by intracellular calcium levels.
The precise mechanism is: (i) calcium binds to the protein calmodulin forming the complex CaM where 4 ions of Ca$^{2+}$ are fixed to calmodulin; (ii) CaM binds to a SK channel terminal to open it; (iii) 4 terminals must be open to let the SK channel open. We now model these different steps.

\paragraph{Saturated calmodulin production.}  The set of kinetic equations leading to CaM formation is widely described in M. Graupner's work \cite{graupner:03,graupner-erler-etal:05}. This is a cascade of equations that we summarize in one kinetic equation, from free calmodulin, $M$, to the saturated one, $CaM$.\\
Let us call $k_{ass} \, (M^{-4} s^{-1})$ and $k_{diss} \, (s^{-1})$ respectively association and dissociation constants of calmodulin. Set $K_d^4 = \frac{k_{diss}}{k_{ass}}$. If we call $M_0 = \left[M\right] + \left[CaM\right]$ the total calmodulin concentration, and $S=\frac{\left[CaM\right]}{M_0}$ the fraction of saturated calmodulin, we have $\frac{\left[M\right]}{M_0}=1-S$ and we obtain a kinetic equation:
$$
\frac{dS}{dt} = k_{ass} C^4 (1-S) - k_{diss} S.
$$
where $C$ is the intracellular calcium concentration.
Setting $\tau_S=\frac{1}{k_{diss}}$ and $\alpha_S=\frac{k_{ass}}{k_{dis}}=\frac{1}{K_d^4}$
we arrive at:
\begin{equation}\label{eq:SApp}
\tau_S \frac{dS}{dt} = \alpha_S C^4 (1-S) - S.
\end{equation}

Note that equation \eqref{eq:S} has a similar form to the one proposed by Hennig et al. in \cite{hennig-adams-etal:09}. However,  in their model our term $C^4$ is replaced by $\frac{C^4}{K_{d}^4 + C^4}$. \\

\paragraph{Binding of calmodulin to SK terminals.} This corresponds to a reaction:
$$
F + CaM 
\begin{array}{ccc}
&\begin{array}{ccc}
&P_{FB}\\
&\rightarrow
\end{array}\\
&\begin{array}{ccc}
&\leftarrow\\
&P_{BF}
\end{array}\\
\end{array}
B,
$$
where $F$ is the density of free terminals, $B$ the density of bounded terminals, $P_{FB}$ ($P_{BF}$) the transition rate from free to bound (bound to free).

Calling $R$ the fraction of bounded terminals, $\tau_{R}=\frac{1}{P_{BF}}$, $\alpha_R = \frac{P_{FB}}{P_{BF}}$ we end up with a kinetic equation:
\begin{equation} \label{eq:RApp}
\tau_{R}\frac{d R}{d t}= \alpha_R \, S(1-R)-R = \alpha_R \, S - (1+\alpha_R \, S)R.
\end{equation}
Finally, since $R$ is the probability that a terminal is open and since $4$ terminals must be open to let the SK channel open, the sAHP conductance is $g_{sAHP} R^4$. 

Note that equation \eqref{eq:R}, is similar to Eq (3) in \cite{hennig-adams-etal:09} ($\tau_{R}\frac{d R}{d t}= (\alpha C+ S)(1-R)-R $), with a remarkable difference: in our model there is no direct dependence on calcium concentration whereas the term $\alpha C$ in \cite{hennig-adams-etal:09} corresponds to a direct binding of Ca$^{2+}$ to a terminal. Note that, taking the quite large value of the parameter $\alpha$ ($\sim 2400$) considered by these authors, their equation is essentially equivalent to $\tau_{R}\frac{d R}{d t}= \alpha \, C (1-R) -R = \alpha \, C  - (1  + \alpha \, C) R $ with a steady state $R=\frac{\alpha \, C}{1  + \alpha \, C}$ very close to $1$ whenever $\alpha \, C$ is quite larger than $1$.  In this case, $S$ plays essentially no role.\\

\paragraph{Calcium concentration.}
Both variables $R$ and $S$ are driven by intracellular Ca$^{2+}$ concentration dynamics, given by:
\begin{equation}\label{eq:Correc_Ca}
\tau_{C}\frac{d C}{d t}=-\frac{\alpha_{C}}{H_{X}}C+C_{0}  + \delta_{C} \, I_{C}(V)
\end{equation}
Equation \eqref{eq:Correc_Ca} is a linear approximation of  a more complex equation (\eqref{eq:Correc_Ca_App} below). This equation is similar to Hennig et al. (Eq (5)) with two significant differences: (1) We have added a rest concentration $C_0$ avoiding unphysical situations where $C$ can become negative; (2) the value of parameters are different. 
\paragraph{Calcium concentration dynamics.}
 The calcium current crossing a membrane section results from the opening of gates in ionic channels. Following \cite{graupner:03} the equation for $Ca$ concentration is (adapted with our notations):
\begin{equation}\label{eq:Gen_Ca}
\frac{d C}{dt} = \frac{G}{n_{Ca} F} \bra{\frac {I_{C}}{S} - J_X(C) - J_p (C) + L} \frac{1}{1 + \frac{d Ca_{bound}}{d C}}.
\end{equation}
Here $ G = \frac{S}{V} = 6 \, \mu m^{-1} = 6 \times 10^{5} \, dm^{-1}$ is the surface to volume ratio that accounts for the localization of the channels at the surface of the membrane, $n_{Ca}=2$ is the calcium valence and $F=96500 \, C  \, mol^{-1}$ is the Faraday number. $\frac{d Ca_{bound}}{d C}$ corresponds  to a 
quasi-steady-state approximation for the calcium buffering where the bound
calcium concentration (on calmodulin) $Ca_{bound}$ is adapted instantaneously to the free calcium concentration $C$ at each time. Since we have no way to estimate $\frac{d Ca_{bound}}{d C}$ we shall consider it is a constant and set $\frac{1}{1 + \frac{d Ca_{bound}}{d C}} \equiv K_{bound}$. To alleviate notations we set:
\begin{equation}\label{eq:r_phys}
r = \frac{G \,K_{bound}}{n_{Ca} \, F }. 
\end{equation}

The first term in Eq \eqref{eq:Gen_Ca}, $r \frac {I_{C}}{S}$ corresponds to an increase of internal $Ca^{2+}$ concentration upon calcium influx (current $I_{C}(V)$) generated by spikes or, in experiments, by voltage clamp. As $V_{Ca}=50$ mV this current is positive unless $V > V_{Ca}$.

The second term is $-r J_X(C)$, where
$$
J_X(C)= \rho_X I_X \frac{C}{H_X+C},
$$
is the efflux current density through sodium-calcium exchanger (NCX). It corresponds to an outward current through NCX exchangers, contributing to restoring the initial $Ca^{2+}$ concentration. 
 Here, $\rho_X$ is the surface density of the NCX membrane proteins. We take $\rho_X= 100 \, \mu m ^{-2}= 10^{12} \, dm^{-2}$ (from \cite{graupner-erler-etal:05}). $I_X= 4.8 \times 10^{-19} \, C \, ms^{-1} = 4.8 \times 10^{-16} \, C \, s^{-1}$ is the 
maximum ionic flux through a single NCX channel. This corresponds to $3$ + charges ($1.5 \, Ca^{2+}$) per $ms$; $H_X=1.8 \mu M = 1.8 \times 10^{-6} M$ is the half activation concentration \cite{graupner:03}. 

Also, in \eqref{eq:Gen_Ca}, $J_p(C)$
%
%
is the current density of $Ca^{2+}$ pumps.
We shall neglect this term from now on.
Finally, $L$ is the leakage surface
current density representing the residual conductivity of the plasma membrane.
We have not been able to find its value in the literature.\\
To summarize Eq \eqref{eq:Gen_Ca} becomes:
\begin{equation}\label{eq:Correc_Ca_App}
\frac{d C}{dt} = r \bra{ \frac {I_{C}}{S} - \rho_X I_X \frac{C}{H_X+C} + L}.
\end{equation}
We also assume that NCX current term is approximated by a linear term. This is valid if one assumes that calcium concentration $C \ll H_X$ which is the case in our simulations. 


In order to match \eqref{eq:Correc_Ca_App} with the form \eqref{eq:Correc_Ca} (\eqref{eq:Ca} in the text),  
%
%
%
%
we set:
\begin{eqnarray}
\frac{\alpha_{Ca}}{\tau_{Ca}} &=& r \rho_X I_X 
; \label{eq:alpha_Ca_exp}\\
\frac{\delta_{Ca}}{\tau_{Ca}} &=&\frac{ r}{S} 
; \label{eq:delta_Ca_exp}\\
\frac{C_0}{\tau_{Ca}}&=& r L  
\label{eq:C0_exp}.\\
\end{eqnarray}


\subsection*{How does the average interburst depend on the parameter  $I_{ext}$.} \label{App:CurvetauVL}

In this section we analyze how the interburst interval $\tau_{IBI}$ depends on  the current $I_{ext}$. In particular, we justify Eq \eqref{eq:tauIBI}. 

As explained in the text there are  two distinct regimes where the interburst interval $\tau_{IBI}$ is constrained by different factors. 
In the noise driven bursting regime ($I_{ext} < I_{SN_1}$), 
the stochastic dynamics around the rest state is described, with a good approximation, by a Ornstein-Uhlenbeck process (upon neglecting the non linearities coming from $g_C$ and $g_K$, which are small in this range of membrane potential). Therefore, fluctuations of $V$ are Gaussian with a mean $V_{rest}$ and a variance proportional to $\sigma$, the noise intensity. Thus, the probability to cross the bifurcation threshold and enter into the bursting regime can be easily computed. This is a sigmoid with a slope proportional to $\frac{1}{\sigma}$.  As a consequence, bursting can take place when $I_{ext} < I_{SN_1}$, because of noise, but the probability to burst decreases rapidly when $I_{ext}$ decreases, after a threshold value $I_c \equiv I_c(\sigma)$ depending on $\sigma$. Formally, the bifurcation value $I_{SN_1}$ is shifted to $I_c$, explaining this term  in Eq \eqref{eq:tauIBI}. For $I_{ext} < I_c$ the probability to induce bursting by noise is so small that we consider it vanishes. Here $\tau_{IBI}$ is not defined anymore but we set it to zero in Eq \eqref{eq:tauIBI} by convention.

In the dynamically driven bursting regime ($I_{ext} > I_{SN_1}$), $\tau_{IBI}$ is constrained by two times: (i) the time $t_1$ to get in a neighborhood of the bifurcation point $V_{SN_1}$ in Fig \ref{Fig:SpontaneousBursting} where bursting starts  (essentially determined by the sAHP) and (ii) the time $t_2$ to leave this neighborhood. When $I_{ext}$ is quite larger than $I_{SN_1}$, $\tau_{IBI}$ is largely dominated by $t_1$. Here, $\tau_{IBI}$ decreases slowly as $I_{ext}$ increases. Indeed, the higher $I_{ext}$ the more neurons are prone to bursting. When $I_{ext}$ becomes too large, $I_{sAHP}$ is not enough to stop the oscillatory phase and $\tau_{IBI}=0$. 

 Close to $I_{SN_1}$, we can use the normal form of the saddle-node bifurcation for the variable $V$: $C \frac{dV}{dt}=-(I_{SN_1}-I_{ext})+ \alpha (V-V_{s})^2$ where  $V_s$ is the rest state at the bifurcation point and $\alpha$ depends on the other parameters of the model. In this approximation, there are two branches of fixed points, existing only when $I_{ext} < I_{SN_1}$, approximated  by $V_{\pm} \sim V_s \pm  \sqrt{ \alpha \, \pare{I_{SN_1}-I_{ext}}}$. The linear stability of the stable branch
is given by the linearized equation $\frac{d x}{dt}=-2  x \, \sqrt{\alpha \pare{I_{SN_1}-I_{ext}}} $ where $x$ is a small perturbation around the fixed point. This equation has a characteristic time $\frac{1}{2 \sqrt{\alpha \pare{I_{SN_1}-I_{ext}}}}$ which diverges as $I_{ext} \to I_{SN_1}$ from below. When $I_{ext} \gtrapprox I_{SN_1}$ the normal form still holds but there is no fixed point anymore. Nevertheless, around the
point where bursting starts the flow has a very small amplitude. As a consequence, the time to leave a neighborhood of this point is also of order $\frac{1}{2 \sqrt{\alpha \pare{I_{SN_1}-I_{ext}}}}$. In this case $\tau_{IBI}$ is largely dominated by the time $t_2$ to leave the neighborhood of the transition point where bursting starts. In the absence of noise, we have thus $\tau_{IBI} = \frac{K}{\sqrt{I_{ext}-I_{SN_1}}}$ for $I_{ext} > I_{SN_1}$, where $K=\frac{1}{2 \sqrt{\alpha}}$. In the presence of noise the threshold value $I_{SN_1}$ is shifted to $ I_c$ as explained above. This fully justifies Eq \eqref{eq:tauIBI}.

\subsection*{Rescaled equations and Multi-time scale analysis}\label{App:Rescaled-Eq} 
The dynamical system \eqref{eq:Voltage1}, \eqref{eq:N}, \eqref{eq:R}, \eqref{eq:S}, \eqref{eq:Ca} has $3$ characteristic time scales: fast variables $V,N$ (of order $ms$); medium $C$ (of order $s$) ; slow $R,S$ (of order $10$ s), fixed by the characteristic times given in the Table \ref{Tab:Dim}. In order to make explicit these time-scales separation we set  $\t{g}_X = \frac{g_X}{g_L}$ for conductances ($X=C, K, sAHP$); $\tau_L = \frac{C_m}{g_L}$, $\tilde{I}_{ext}=\frac{I_{ext}}{g_L}$; $t_f=\frac{t}{\tau_L}$; $\t{\tau}_X=\frac{\tau_X}{\tau_L}$, where $X=N,C,R,S$. 

This gives:
$$
\left\{
\begin{array}{lllll}
\frac{d V}{d t_f}&=&-(V-V_L)-\t{g}_C M_\infty(V)(V-V_C)-\t{g}_K N(V-V_K) - \t{g}_{sAHP}R^4(V-V_K) + \tilde{I}_{ext}\\
&&\\
\tilde{\tau}_N\frac{d N}{d t_f}&=&\Lambda(V)(N_{\infty}(V)-N)\\
&&\\
\tilde{\tau}_C\frac{d C}{d t_f}&=&-\frac{H_X}{\alpha_C} C+C_0 - \delta_C g_C M_\infty(V)(V-V_C)\\
&&\\
\tilde{\tau}_{S}\frac{d S}{d t_f}&=&\alpha_{S}(1-S) C^4 - S\\
&&\\
\tilde{\tau}_{R}\frac{d R}{d t_f}&=&\alpha_{R} S(1-R) - R\\
\end{array}
\right.
$$

Note that, with our parameters values, $\tau_L=11$ ms, $\frac{1}{\tilde{\tau}_C} \sim 5.5 \times 10^{-3}$, $\frac{1}{\tilde{\tau}_S} =  \frac{1}{\tilde{\tau}_R} \sim 2.5 \times 10^{-4}$. Therefore, on the fast time scale, one can use the approximation $\frac{1}{\tilde{\tau}_X} =0$, $X=C,S,R$, so that the variables ${C},{S},{R}$ are constant. So, fast dynamics reduces to a Morris-Lecar model (here with a fast variable $N$) in the presence of an additional current $I_{tot} =  I_{sAHP} + I_{ext}$:
\begin{equation}\label{eq:ML_fast}
\left\{
\begin{array}{lll}
C_m \frac{d {V}}{d {t}}&=&-g_L (V-V_L)-{g}_{C} {M}_{\infty}({V})({V}-{V}_{C})-{g}_{K}{N}({V}-{V}_{K}) + I_{tot};\\
&&\\
{\tau}_N\frac{d {N}}{d {t}}&=&{\Lambda}({V})({N}_{\infty}({V})-{N}).
\end{array}
\right.
\end{equation}

\subsection*{Parameters values and auxiliary functions} \label{App:Parameters_Value}

\paragraph{Units.} In all the paper, physical quantities are expressed in the units displayed in Table \ref{Tab:Dim}. Having integrated over all the surface of the membrane we omit the surface units. 

\paragraph{Calibrating parameters from experiments}
All parameters values are calibrated with respect to biophysics, found in the literature or fitted from experimental curves in \cite{abel-lee-etal:04}, \cite{zheng-lee-etal:04} and \cite{zheng-lee-etal:06}. 
Morris-lecar tuning parameters $V_1$, $V_2$, $V_3$, $V_4$ were calibrated (see Fig \ref{fig:Zhou-Kmodel}), so as to reproduce the experiment of \cite{zheng-lee-etal:06} (Fig 4a), where the authors investigate the ionic mechanisms of the fast oscillations. Note that the bursting regime is robust to (small) variations of these parameters (results not shown).
We tuned the sAHP parameters taking into account the analogy with SK channels studied in \cite{abel-lee-etal:04} (fit not shown). Also, we note that the intensity of sAHP observed by Abel et al. in pyramidal neurons (of order $150$ pA) is quite bigger than in stage II SAC. In our model, this means a lower sAHP conductance $g_{sAHP}$ ($g_{sAHP}=2$ nS).

\begin{table}[htb!]
\begin{center}
{\bf \caption{Units of physical quantities used in the paper. \label{Tab:Dim}}}
\begin{tabular}{|c|c|}
   \hline
   Physical quantity & Units  \\    \hline
	Time & ms \\ \hline
	Potential & mV  \\ \hline
	Capacitance & pF \\ \hline
	Current & pA \\ \hline
	Conductance & nS \\ \hline
	Concentrations & nM \\ \hline
 \end{tabular}

 \end{center}
\end{table}

\paragraph{Auxiliary functions.} The dimensionless auxiliary functions involved in the dynamical equations appearing in the model definition are:

\begin{equation}\label{eq:gcav}
M_\infty(V)=\frac{1}{2} [1+\tanh (\frac{V-V_{1}}{V_{2}})],
\end{equation}

\begin{equation} \label{eq:Lambda}
\Lambda(V)=\cosh(\frac{V-V_{3}}{2 V_{4}}),
\end{equation}

\begin{equation}\label{eq:Ninf}
N_{\infty}(V)=\frac{1}{2} [1+\tanh (\frac{V-V_{3}}{V_{4}})],
\end{equation}

\paragraph{Parameters.} The parameters used in the model are displayed in Table \ref{Tab:Param}.

\begin{table}[!htt]
\begin{center}
    \begin{tabular}{ | p{2cm} | p{2.5cm} | }     \hline
    Parameter & Physical value \\ \hline
    {{$C_m$}} & $22\, pF$  \\ \hline
    {{$g_L$ }}& $2\, nS$  \\  \hline
    {{$g_{C}$}} & $[3,20] \, nS$\\  \hline
    {{$g_{K}$}} & $[1,20]\, nS$\\ \hline   
   	{{$g_{sAHP}$}} & $2\, nS$ \\ \hline    
    {{$V_L$}}  & $-70 \, mV$\\ \hline
    {{$V_{C}$}}  & $50 \, mV$ \\ \hline
    {{$V_K$}}  & $-90  \, mV$   \\ \hline
    {{$V_1$}}  & $-20  \, mV$  \\ \hline
    {{$V_2$}}  & $20 \, mV$ \\ \hline
    {{$V_3$}}  & $-25 \, mV$\\ \hline
    {{$V_4$}}  & $7 \, mV$ \\ \hline
    {{$\tau_{N}$}}  & $5 \, ms$    \\ \hline
    {{$\tau_{R}$}}  & $8300 \, ms$ \\ \hline
    {{$\tau_{S}$}}  & $8300 \, ms$  \\ \hline
    {{$\tau_{C}$}}  & $2000 \, ms$  \\ \hline
    {{$\delta_{C}$}}  & $10.503 $ $nM \,\, pA^{-1}$ \\ \hline
    {{$\alpha_S$ }}  & $\frac{1}{200^4}$ $nM^{-4}$  \\ \hline
    {{$\alpha_{C}$}}  & $4865$\, nM  \\ \hline
    {{$\alpha_{R}$}}  & $4.25$  \\ \hline
     {{$H_X$}}  & $1800$ $nM$  \\ \hline
    {{$C_0$}}  & $88$ $nM$ \\ \hline
   \end{tabular}
\end{center}
\caption{Range of values for the parameters used in the paper.}
\label{Tab:Param}
\end{table}

\subsection*{Comparison with existing models\label{comparemodels}} 

In this section we shortly revisit models of SACs activity in the stage II and compare them to our model, (see \cite{burgi-grzywacz:94,butts-feller-etal:99,godfrey-swindale:07,hennig-adams-etal:09,ford-felix-et-al:12,lansdell-ford-etal:14}, for a review see \cite{godfrey-eglen:09}). We would first like to remark that these existing models are devoted
to describe wave activity and do not focus on thoroughly describing individual SACs 
dynamics. Especially, none of the models we know describe the biophysical mechanisms of SACs bursting activity and the role played by biophysical parameters. Instead their focus was more on having a relatively simple description of the cell activity with a minimal set of tunable parameters (a notable example is Butts et al. model \cite{butts-feller-etal:99} which has two free parameters governing the waves properties). 
Finally, our model is the first one, able to characterize SACs dynamics in great detail and to show that biophysical properties of SACs are sufficient to explain intrinsic bursting and refractoriness in these cells.

The closest model to ours has been proposed by M. Hennig and collaborators (refered to as "Hennig model" \cite{hennig-adams-etal:09}, see also the extension by Ford and Feller and the recent paper of Xu et al. \cite{xu-burbridge-etal:16}). Actually, our model has been widely inspired by this work with several notable differences. As exposed in the Methods section our biophysical analysis of sAHP dynamics leads to equations and parameters values departing from Hennig model. Additionally, Hennig model does not consider a fast potassium dynamics
and there is no fast oscillation. The mechanism that mimics SACs bursting is a switch from low membrane potential level to high one. This switch is determined by an exogenous shot noise i.e. a voltage dependent rate modulated Poisson process with a slow decay. This activity is maintained long enough so that sAHP can be activated, enabling the cell to return to rest. In our model shot noise is not necessary to trigger activity. Instead, a cell can spontaneously switch to the bursting state, where it stays until the sAHP produced by its activity leads it back to the rest state. By “spontaneous” we mean literally "happening or done in a sudden way, without any planning or without being forced/without premeditation". In our model this sudden switch is a bifurcation, induced by the mere cells dynamics. The presence of a fast (Brownian) noise facilitates this transition, but, there is no need for a shot noise. The cell stays in the bursting state by its mere dynamics, even when it is isolated.

A similar modeling to Hennig holds in Lansdell et al. model \cite{lansdell-ford-etal:14}.
It is ruled as well by an excitable Morris-Lecar model with a slow potassium variable linked to sAHP. There is no fast potassium and here too cells do not burst. As in Hennig model, the cell activity is triggered by a random excitatory current and maintained by network dynamics.

Although Hennig or Lansdell model are based on differential equations with many parameters, none of these authors made a bifurcation analysis of their model. We did it and we found that there is no bifurcation in a neighborhood of the parameters value they choose.
In this sense, our model is in strong contrast with previous studies. \textit{This is precisely because cells are close to a bifurcation point that they are able to exhibit the wide repertoire of dynamics we have presented in close agreement with experimental findings.}

%
%
%

\section*{Author Contribution statement.}
D.K. contributed to the model definition-analysis and to the writing of the paper, she made numerical simulations and bifurcations studies. 
L. G. contributed to the model definition and to the writing of the paper.
E. O. contributed to the biological interpretation of the model.
O. M. contributed to the model definition and to its biological interpretation.
S. P. contributed to the model definition and to its biological interpretation.
B. C. contributed to the model definition and analysis and to the writing of the paper, he made numerical simulations and mathematical developments. He supervised the research and paper redaction.

\section*{Competing interests}
The authors declare no competing interests.

\section*{Acknowledgements}
D.K. was supported by a doctoral fellowship from Ecole Doctorale des  Sciences et Technologies de l'Information et de la Communication de Nice-Sophia-Antipolis (EDSTIC).  E.O. was supported by a doctoral fellowship from Ecole des Neurosciences de Paris Ile-de-France. This work also benefited from the support of the axis MTC-NSC of the University Côte d'Azur and the Doeblin federation. We warmly acknowledge Matthias Hennig and Evelyne Sernagor for their invaluable help. We thank Pr. Z. Jimmy Zhou for his authorization to reproduce Fig \ref{fig:Zhou-Kmodel} A, and \ref{Fig:P4} A. We thank the reviewers for their helpful criticism. 

\end{document}